\documentclass[12pt,preprint]{aastex}
\begin{document}

\title{The chemical composition of red giant stars in four 
intermediate-age clusters of the 
Large Magellanic Cloud
\footnote{Based on observations obtained at Paranal ESO Observatory 
under proposal 072.D-0337(A), 072.D-0342(A) and 074.D-0369(A).}}

\author{Alessio Mucciarelli}
\affil{Dipartimento di Astronomia, Universit\`a 
degli Studi di Bologna, Via Ranzani, 1 - 40127
Bologna, ITALY}
\email{alessio.mucciarelli@studio.unibo.it}

\author{Eugenio Carretta }
\affil{INAF - Osservatorio Astronomico di Bologna, Via Ranzani, 1 - 40127
Bologna, ITALY}
\email{eugenio.carretta@oabo.inaf.it}

\author{Livia Origlia}
\affil{INAF - Osservatorio Astronomico di Bologna, Via Ranzani, 1 - 40127
Bologna, ITALY}
\email{livia.origlia@oabo.inaf.it}

\author{Francesco R. Ferraro}
\affil{Dipartimento di Astronomia, Universit\`a 
degli Studi di Bologna, Via Ranzani, 1 - 40127
Bologna, ITALY}
\email{francesco.ferraro3@unibo.it}

\begin{abstract}

This paper presents the chemical abundance analysis of a sample of 27 
red giant stars located in 4 popolous intermediate-age globular 
clusters in the Large Magellanic Cloud, 
namely NGC 1651, 1783, 1978 and 2173. This analysis is based 
on high-resolution (R$\sim$47000) spectra obtained with the 
UVES@VLT spectrograph. 
For each cluster we derived up to 20 abundance ratios sampling 
the main chemical elemental groups, namely light odd-Z, 
$\alpha$, iron-peak and neutron-capture elements. \\
All the analysed abundance patterns behave similarly in the 4 clusters and 
also show negligible star-to-star scatter within each cluster.
We find [Fe/H]=-0.30$\pm$0.03, -0.35$\pm$0.02, -0.38$\pm$0.02 and 
-0.51$\pm$0.03 dex for NGC 1651, 1783, 1978 and 2173, respectively.\\
The measurement of light odd-Z nuclei gives slightly subsolar [Na/Fe] and
a more significant [Al/Fe] depletion ($\sim$-0.50 dex). 
The [$\alpha$/Fe] abundance ratios are nearly solar, while
the iron-peak elements well trace that one of the iron.\\
s-process elements behave in a peculiar way:
light s-elements give subsolar [Y/Fe] and [Zr/Fe] abundance ratios, 
while heavy s-elements give 
enhanced [Ba/Fe], [La/Fe] and [Nd/Fe] with respect to the solar values. 
Also, the [Eu/Fe] abundance ratio turns out to be enhanced ($\sim$0.4 dex).

\end{abstract}


\section{Introduction}

The Large Magellanic Cloud (LMC) is the nearest galaxy with a present-day 
star-formation activity and it represents a formidable laboratory 
for the study of stellar populations.
Its globular cluster (GC) system
shows a wide distribution of ages \citep{swb,ef88,geisler97}, 
metallicities \citep{sp89,ols91} and integrated colors \citep{vdb81,pers83}. 
In particular, we can distinguish three main stellar populations:
an old and metal poor population \citep[$\sim$13 Gyr, see e.g.][]{broc,olsen}, 
the analogous of the Galactic halo GCs, an intermediate-age population 
\citep[$\sim$1-3 Gyr,][]{gallart, f04, m06a, m07} and 
a young population, with clusters younger than 1 Gyr \citep{fischer,broc03}.
The lack of objects with ages in the 
$\approx$3-10 Gyr range, the so-called {\sl Age Gap} 
\citep{rich01,bekki,mac06} 
represents one of the long-standing
problems related to the clusters formation history of the LMC 
(a similar {\sl Age Gap} is not seen in the Small Magellanic 
Cloud (SMC) clusters). \citet{bekki_06} discussed three possible 
scenarios to solve this problem: (1) after the initial burst
of clusters at the epoch of the galaxy formation ($\sim$13 Gyr ago) 
the cluster formation has been interrupted until $\sim$3 Gyr ago; 
(2) the cluster formation has been not suspended after the initial 
burst. The cluster with ages between $\sim$13 and 3 Gyr have been 
tidally stripped, or 
(3) preferentially destroyed by the LMC tidal field. 
The most recent theoretical investigations \citep{bekki,bekki_06} 
have shown that the main episodes of star formation in the LMC 
can be related to the close encounters with the SMC. 
These latter events could be also responsible for the formation 
of the off-center Bar and the age distribution of the LMC 
GC system. In particular, the first, very close 
encounter between the LMC and SMC ($\sim$4 Gyr ago) was able to 
re-ignite the cluster formation, with a rapid chemical 
enrichment due to this very efficient star formation 
activity \citep[see also ][]{pt98}. Moreover, a possible 
infall of metal-poor material from the SMC could be 
the origin of some metal-poor, young LMC clusters, 
observed by \citet{ols91} and \citet{aaron}.

The LMC GCs are ideal tracers of the chemical evolution 
of their host galaxy, recording in their abundance patterns 
the level of enrichment 
in the galactic environment at the time of their formation. 
However, our knowledge of the chemical abundances of the LMC 
GCs is still very sparse and uncertain.
Most of the information still 
rely on photometry \citep{jt94, pacheco, dirsch}
or low-resolution spectroscopy (see e.g., \citet{ols91} and \citet{aaron}).
Recently, \citet{aaron} derived a new, 
homogeneous metallicity scale for 23 intermediate-age and 5 
old LMC clusters based on the Ca II triplet lines, 
observed with FORS2@VLT. 
The derived mean metallicities are  
[Fe/H]\footnote{We adopt the usual spectroscopic notation: 
[A]=log$(A)_{star}$-log$(A)_{\odot}$ for each
element abundance A; log(A) is the abundance by number of the 
element A in the standard scale 
where log(H)=12.}=-0.48 dex (with rms=0.09) and 
[Fe/H]=-1.66 dex (with rms=0.27) for the 
intermediate-age and old clusters, respectively. 

Despite the new generation of 8-meter class telescopes, 
detailed chemical information 
about the LMC clusters from high-resolution spectra are limited to 
a few stars in a few clusters and they are insufficient to draw 
a global picture of the chemical 
properties of these objects and to constrain the timescales of the 
chemical enrichment.\\
Several studies have concerned 3 young, populous clusters, namely NGC 1818 
\citep{richtler, korn00}, NGC 2004 \citep{korn00, korn02} 
and NGC 2203 \citep{smith02}, indicating a high metallicity 
([Fe/H]$>$-0.6 dex) and a mild deficiency of $\alpha$-elements.
\citet{hill00} presented chemical abundances of Fe, O and Al  
from high-resolution spectra of 10 red giant stars in 4  
LMC globular clusters, namely NGC 1866, NGC 1978, ESO 121 and NGC 2210,
spanning the entire age range of the LMC clusters system.
They found 
[Fe/H]=-0.50, -0.96, -0.91 and -1.75 dex, respectively and 
slightly enhanced [O/Fe] and slightly depleted [Al/Fe] with 
respect to the solar ratios.
\citet{johnson06} presented detailed abundances for 10 giant stars in 
4 old globular clusters, namely NGC 1898, NGC 2005, NGC 2019 and Hodge 11, 
finding [Fe/H]= -1.23, -1.47, -1.37 and -2.21 dex, respectively.
They generally found abundance ratios comparable to 
those of the Galactic GCs. Exceptions are the [Ca/Fe] and 
[Ti/Fe] ratios, similar to the solar values, and
[V/Fe] and [Ni/Fe] which are significantly underabundant 
 (by a factor of 2-3) with respect to the solar ratio.
Finally, \citet{f06} reported the iron content of the 
intermediate-age cluster NGC 1978 based on the analysis of eleven 
giant stars observed with the high resolution 
spectrograph FLAMES@VLT. For this cluster a large metallicity 
dispersion has been claimed by \citet{hill00} based on 2 stars only,
but \citet{f06} new analysis shows no significant dispersion 
of the iron content.

On the other hand, several works have been addressed to investigate 
the metallicity of the LMC field. Abundances from mid-high 
resolution spectroscopy are available for few samples of giant 
and supergiant stars \citep{rb89,mcw91,hill95,smith02} 
and variables \citep{luck}. 
Furthermore, \cite{cole05} presented a spectroscopical survey 
for 373 red giants in the LMC
bar, based on Ca II triplet analysis. The derived metallicity
distribution function is peaked at the median value of 
[Fe/H]=-0.37 dex and only a small number of stars shows a 
metallicity [Fe/H]$<$-2.10 dex. \\
\citet{pompeia06} reported abundance ratios for 62 giant stars 
located in the inner disk of the LMC, finding an average 
[Fe/H]=--0.80 dex (rms=0.29 dex), 
roughly solar $[\alpha/Fe]$ and [iron-peak/Fe] abundance ratios.
They also measured s-process elements, finding a depletion
of [light-s/Fe] and an enhancement of [heavy-s/Fe] with respect to solar. 
Very recently, \citet{carrera07} analysed $\sim$500 giant stars 
at different distance from the LMC center; their analysis, 
based on Ca II triplet, confirms the previous surveys, with a 
mean metallicity constant until to $6^{\circ}$ from the 
LMC center ([Fe/H]$\sim$-0.5 dex) and a weak decrease in 
the outermost field ([Fe/H]$\sim$-0.8 dex).

The present study is the second \citep[after][]{f06} of 
a long-term project devoted to obtain a complete screening 
of the chemical abundances and abundance patterns of a 
sample of pillar LMC GCs. In this paper we describe the 
chemical abundance analysis of 27 red giant stars members of 4 
LMC clusters, belonging to the intermediate-age population. 
Future 
papers will be devoted to study the chemical properties 
of the young and old cluster populations.
The overall goal of this project is twofold:\\ 
(1) the definition of 
a new and homogeneous 
metallicity scale for the LMC GC system based on high 
resolution spectra of giant stars, members of a representative 
number of {\sl pillar} clusters, sampling 
different ages. 
This scale, combined with high-quality optical 
photometric datasets, will be crucial to obtain precise ages 
for these clusters \citep[see the recent results in][]{m07};\\ 
(2) a detailed comparison of the cluster populations and their 
chemical abundance patterns with those in the LMC fields and 
in other galactic environments. This is crucial to constrain 
the age-metallicity relation and the overall star formation 
and chemical enrichment of the LMC.
 
In particular, the chemical signatures of the 
intermediate-age LMC GCs (a class of objects still poorly 
studied, especially at high spectral resolution) 
provide information
about the global metallicity of the LMC (these clusters represent 
the majority of the entire LMC cluster population, as discussed by 
\citet{ols91}) and to test the level of chemical enrichment 
in the last 3 Gyr, after the restart of the cluster formation, 
according to the scenario drawn by \citet{bekki_06}. Moreover, 
a detailed knowledge of the whole chemistry of the GC system 
(both young and old clusters) is fundamental to understand 
the formation of the dwarf irregulars (like LMC and SMC) 
in the framework of the hierarchical models 
\citep[see the accurate review by][]{geisler}.

This paper is organized as follows: Sect.~2 describes the 
observational dataset, Sect.~3 the computation of the atmospheric parameters, 
the measurements of the line equivalent widths (EW) and the error budget. Sect.~4 
presents the results of our abundance analysis for the most important chemical elements.
Finally, Sect.~5 and 6 report our discussion and 
conclusions.

\section{Observational data}

The observations were performed by using the multi-object spectrograph FLAMES 
\citep{pasquini02}, mounted at the Kueyen 8 m-telescope (UT2) of the ESO Very 
Large Telescope on Cerro Paranal (Chile).
We used FLAMES in the UVES+GIRAFFE/MEDUSA combined mode for a total of 
8 UVES and 132 MEDUSA fibres. 
Here we present the results of the UVES Red Arm survey which provides 
high resolution (R$\sim47000$) spectra in the 4800-6800 $\mathring{A}$ 
wavelength range of 6-7 stars in one shot.
The spectra were acquired during 3 nights allocated 
to the ESO Program 072.D-0342(A).
Additional observations were performed as back-up programmes 
in two Visitors Mode runs (ESO Program 072.D-0337(A) and 
ESO Program 074.D-0369(A)).
The selection of the target stars is based on our high quality near-infrared 
(J, H and K filters) photometric catalogs of a large sample of LMC clusters, as 
secured by our group \citep{f04, m06a}.
These catalogs have been astrometrized onto the 2MASS system.  
The selected stars for the spectroscopic survey belong
to the brightest portion of the RGB (K$<$14), whose tip is located at 
$\rm K_{0}\approx$12.1 \citep{cioni}, in order to minimize the 
possible contamination by AGB stars.
Fig. 1 shows the (K, J-K) CMDs of the 4 clusters
with marked the spectroscopic targets. \\
The spectra have been acquired in series of 4-6 exposures of $\approx$45min each: 
the pre-reduction procedure has been performed by 
using the UVES ESO-MIDAS pipeline \citep{mulas2002}, 
which includes bias subtraction, flat-field correction 
and wavelength calibration with a reference Th-Ar calibration lamp. 

All the exposures relative to a given star have been sky-subtracted, corrected 
for radial velocity (by using several tens of metallic lines) and 
average-combined together, providing a final, equivalent spectrum of
 total exposure time of 3-5 hrs, with a 
typical  S/N$\approx$30-40 at about 6000 $\mathring{A}$). 
The radial velocities included heliocentric corrections, calculated by 
using the IRAF task RVCORRECT.
We find  
$\rm v_r=$233.1$\pm$1.8 km/s (rms=3.6 km/s),  
$\rm v_r=$277.6$\pm$1.0 km/s (rms=2.3 km/s),  
$\rm v_r=$236.8$\pm$0.4 km/s (rms=1.2 km/s),  
$\rm v_r=$293.1$\pm$1.5 km/s (rms=3.1 km/s) 
for NGC~1651, NGC~1783, NGC~1978 and NGC~2173, respectively.  
These values 
are in excellent agreement with previous determinations 
by \citet{ols91} and \citet{aaron}.
Tab. 1 lists the main data for each observed star: S/N, 
heliocentric radial velocity, near-infrared magnitude $J_0$ and 
color $(J-K)_0$, right ascension and declination 
\citep[from the catalog by][]{m06a}.

\section{Analysis of the observed spectra}                             

This section describes the analysis of the observed spectra and the 
computation of the stellar parameters,  
providing measurements 
of line EWs, stellar 
temperatures (T$_{eff}$), gravities (log~g) and microturbulence ($v_t$)
and the corresponding error budget.

\subsection{Equivalent widths}

The analysis of the observed spectra and the computation of the 
chemical abundances (for Fe and other elements) was performed by
using the ROSA package \citep{g88}. 
The line EWs from the observed spectra have been measured by 
Gaussian fitting of the line profiles, adopting a relationship 
between EW and FWHM 
\citep[see e.g.][]{brag}.
The local continuum has been derived by applying an iterative 
clipping average over the points with highest counts 
around each line.\\
An empirical estimate of the internal error in the measurement of EWs 
can be obtained by comparing a large sample of line EWs in pairs of 
stars with similar physical parameters. We derived an average rms of 
13.3, 8.5, 10.3 and 10.5 m$\mathring{A}$ for NGC 1651, NGC 1783, NGC 1978 and
 NGC 2173, respectively. 
Such rms estimates should be divided by $\sqrt{2}$, since
they are distributed in equal proportion to the two stars 
in each pair, thus giving final values of
9.4, 6.0, 7.3 and 7.4 m$\mathring{A}$ for the 4 clusters, respectively. 
These errors are larger than those obtained  
from the \citet{cayrel} formula (which yields a typical uncertainty 
of $\approx$4.2  m$\mathring{A}$), since the latter neglects 
for example the 
uncertainty in the continuum location,
which is the dominant source of error in 
metal-rich, crowded spectra of rather cool giants.
Table 2 lists some measured EWs, the complete set being 
available in electronic form.

\subsection{Stellar parameters}

Stellar $T_{eff}$ were obtained 
from the near-infrared color (J-K), corrected for reddening  
by using the E(B-V) values from \citet{pers83}, 
and the extinction law defined by \citet{rl85}. 
We adopted two different color-$T_{eff}$
transformations, namely by \citet{m98} and \citet{alonso99,alonso01}.
Since the derived temperatures are well in agreement 
within $\le$50 K, we used the average 
of the two values.

Gravities were estimated by using the relation 
between $T_{eff}$, stellar mass and luminosity: 
$$log(\frac{g}{{g_{\odot}}})=4log(\frac{T_{eff}}{T_{eff,\odot}})+0.4(M_{bol}-M_{bol,\odot})+log(\frac{M}{M_{\odot}}),$$
by adopting the solar references log~${g_{\odot}}$=4.437, 
$T_{eff,\odot}$=5770 K and $M_{bol,\odot}$=4.75, 
according to the IAU recommendations \citep{ander}. 
For each target star $M_{bol}$ has been estimated from the K magnitude, 
and using $(m-M)_0$=18.5 \citep{vdb98, alves}
and bolometric corrections by \citet{m98}.
Stellar masses have been estimated by using suitable 
isochrones \citep{cast03,car04}, adopting ages derived 
from the s-parameter \citep{ef88, girardi} calibration, 
and an average
metallicity of Z=0.008 \citep{cole05}, typical of the LMC.
All the program stars have similar $T_{eff}$ ($\sim$3600-4000 K), 
and gravities log~g ($\sim$0.5-1.2 dex).

Microturbulent velocities were estimated by eliminating 
the trend of abundances with the expected line strengths, 
accordingly to the prescription of \citet{magain} and by 
using a large number (80-90) of Fe I lines for each star.
 
The model overall metallicity [A/H] was chosen as that of the 
model atmosphere extracted from the grid of ATLAS models 
by \citet{kur}, with the overshooting option switched on, 
whose abundance matches the one derived from  Fe I lines 
when adopting the appropriate atmospheric parameters 
for each star. The atomic parameters, line list and solar 
reference abundances adopted 
in this analysis are described in \citet{gratton03} and \citet{eu04}.

Fig. 2 shows, as an example of consistency check, the iron 
abundance log~n(Fe) from the Fe I lines as a function of the excitation 
potential $\chi$, expected line strength and wavelength for one of 
the star in our sample, together with the linear fit to each 
distribution (dashed lines).  The log~n(Fe)-$\chi$ relationship 
for each target star shows only a marginal slope ($\sim$ -0.02 dex/eV 
and reported in Fig. 3 as a function of the $T_{eff}$), 
confirming that the adopted photometric $T_{eff}$ well 
reproduce the excitation equilibrium. The lack of significant 
trends in the relationship between iron abundance and the 
expected line strength (Fig. 2, middle panel) supports the 
validity of the adopted $v_t$ values.
The derived iron abundances show no trend with the 
wavelength (Fig. 2, lower panel); this represents a good sanity 
check regarding the continuum placement.

\subsection{Error budget}

We computed the total uncertainty in the derived abundances, 
according to the treatment discussed by \citet{mcw95}.
The variance of a generic abundance ratio [X] is estimated by 
using the following formula:
$$\sigma^{2}_{[X]}=\sigma^{2}_{EW}+\sigma^{2}_{T_{eff}}\cdot({{\partial [X]}\over{\partial {T_{eff}}}})^{2}+\sigma^{2}_{\log{g}}
\cdot({{\partial [X]}\over{\partial {\log{g}}}})^{2}+\sigma^{2}_{[A/H]}\cdot({{\partial [X]}\over{\partial {[A/H]}}})^{2}
+\sigma^{2}_{v_t}\cdot({{\partial [X]}\over{\partial {v_t}}})^{2},$$
where $\sigma_{EW}$ is the abundance uncertainty  
due to the error in the EW measurement, $\sigma_{i}$  
is the internal error related to the atmospheric parameter {\it i} 
and $\partial [X]\over{\partial {i}}$ indicates the differential 
variation of the derived abundance [X] with respect to the 
atmospheric parameter {\it i}. 
These latter terms have been computed for all the elements analysed
in this work by re-iterating the analysis varying each time only 
one parameter, by assuming variations of 
$\Delta{T_{eff}}$=100K, $\Delta{log~g}$=0.2 dex, $\Delta[A/H]$=0.1 dex 
and $\Delta{v_t}$=0.2 km/s. Tab. 3 reports 
the results of such an analysis for the star NGC 1783-29. 
The terms of covariance that measure the correlation between the 
atmospheric parameters are not included in the above formula.

The main error sources in the determination of $T_{eff}$ are the 
photometric error related 
to (J-K) color and reddening E(B-V). All the program stars 
are brighter than K$\sim$14 
and the typical photometric error for the (J-K) color is 
$\sim0.03$ mag; 
for the reddening we assumed a conservative uncertainty of 
$\sim20$\%. These terms translate 
into a $\pm$60 K temperature uncertainty.
In computing the uncertainty due to the stellar gravity, 
we took into account four main error 
sources: the error in $T_{eff}$, in mass ($\pm$10\%), 
in distance modulus ($\pm$0.1 mag) and in bolometric correction 
($\pm$0.05 mag). 
From the quadratic sum of these uncertainties, a total error in 
$\log{g}$ of $\pm$0.08 dex has been obtained.
To estimate the error in the microturbolent velocity we repeated 
the analysis by changing the $v_t$ value until the 1$\sigma_{slope}$ 
value for the slope of the abundance - expected line strength relation 
has been reached. The internal error associated to $v_t$ is 
typically 0.10-0.17 km/s.  

An estimate of the error in the derived abundances due to the 
uncertainty in the measurement of EWs has been estimated by 
weighting the average Fe I line-to-line scatter 
(0.18 dex) with the square root of the mean number 
of measured lines $N_i$ for each {\sl i} element: 
$$\sigma_{EW}=\frac{0.18}{\sqrt{N_i}}.$$ 
Finally, we assumed an additional $\pm$0.1 dex uncertainty 
due to the choice of the best-fit model atmosphere.

\section{Chemical abundances} 

Tab. 4-9 report the abundances of all analysed elements (with the number of 
measured spectral lines and the corresponding line-to-line scatter) for the 
target stars and Tab. 10 and 11 summarize the average ratios for the 4 LMC clusters 
with the number of used stars, the observed star-to-star scatter ($\sigma_{obs}$) 
and the expected error ($\sigma_{exp}$) computed accordingly to the procedure 
described in Sec. 3.3.
For each cluster, Fig. 4-8 plot the average values of the derived 
abundance ratios (big grey points). For comparison, the corresponding 
abundance ratios of other intermediate-age stellar populations are 
reported, namely LMC disk giant stars 
by \citet{pompeia06} (empty triangles), intermediate-age LMC cluster 
giants by \citet{hill00}  (empty squares), Galactic thin disk 
dwarfs by \citet{reddy}  (small grey points) and Sgr giant stars 
by \citet{boni00,monaco05,monaco07,sbordone} 
(small black points).

\subsection{ Iron and Iron-peak elements}

The mean iron abundance of the cluster NGC~1651 results 
[Fe/H]=-0.30$\pm$0.03 dex with rms=0.07 dex, whereas \citet{ols91} 
derived [Fe/H]=-0.37$\pm$0.20 dex. 
Photometric determinations have been presented by 
\citet[][from Stromgreen photometry]{dirsch} 
and \citet[][from isochrones fitting]{sar02}, suggesting 
[Fe/H]=-0.65 dex and [Fe/H]=-0.07$\pm$0.10 dex, respectively. 
Recently, \citet{aaron} estimated [Fe/H]=-0.53$\pm$0.03 dex,
by using the Ca II triplet of 9 giant stars.\\ 
NGC~1783 shows a mean iron abundance of [Fe/H]=-0.35$\pm$0.02 
dex with rms=0.06 dex. For this cluster only photometric 
determinations are available : 
\citet{sp89} found [Fe/H]=-0.45 dex and \citet{pacheco} 
found [Fe/H]=-0.75 dex. \\ 
The results about the iron abundance of NGC~1978 
([Fe/H]=-0.38$\pm$0.02 dex with rms=0.07 dex)
have been discussed by \citet{f06}.\\
The iron content of NGC~2173 turns out to be 
[Fe/H]=-0.51$\pm$0.03  with rms=0.07 dex. 
\citet{ols91} give [Fe/H]=-0.24$\pm$0.20 dex,  
\citet{pacheco} found [Fe/H]=-0.50 dex
and 
\citet{aaron} found [Fe/H]=-0.42$\pm$0.03 dex,
by using the Lick index 
and the Ca II triplet, respectively.

We also measured lines of several elements of the Fe-group, 
namely Sc, V, Cr, Co and Ni. 
Corrections for the hyperfine structure (HFS) due to 
non-zero nuclear magnetic moment, were applied to the ScII, 
V and Co lines, as in 
\citet[][and references therein]{gratton03}. 
The abundance ratios between these elements and
Fe is roughly solar in all the 4 clusters. In order to cross-check 
the abundances derived from the EW measurements, we performed 
a synthetic spectrum fitting for some lines of these elements, 
finding a negligible 
difference between these two determinations.

\subsection{Light odd-{\it Z} elements}

Na abundances were derived from the $\lambda \lambda$5682-88 $\mathring{A}$ 
and $\lambda \lambda$6154-60 $\mathring{A}$ 
doublets and they include non-LTE corrections computed accordingly to \citet{gr99}. 
The differences between LTE and non-LTE derived abundances are generally as
large as $\sim$0.2 dex, with a maximum discrepancy of $\sim$0.35 dex in the coolest 
star of the sample. Three clusters (NGC~1651, NGC~1783 and NGC~1978) exhibit 
mild depletion of [Na/Fe]$\le$-0.1 dex while this ratio is solar in NGC~2173, 
without appreciable intrinsic star-to-star scatter. \\  
Al abundances were derived from the $\lambda \lambda$6696-98 $\mathring{A}$ doublet. 
These lines do not include non-LTE corrections, following the extensive 
discussion by \citet{baum}. 
All the target clusters are characterized by a significant depletion of [Al/Fe], 
typically $\le$ -0.3 dex. Also for this ratio, 
the intrinsic star-to-star scatter is negligible. 
NGC 1978 was previously observed by \citet{hill00} who found 
[Al/Fe]=0.10 dex from the analysis of 2 giants, only. 
This value turns out to be $\sim$0.6 dex higher than that found here 
([Al/Fe]=-0.52 dex), the discrepancy is likely due to the 
strong difference ($\sim$0.6 dex) in the iron content derived from the two 
analysis \citep[see][]{f06}. Indeed, their [Al/H] abundance is consistent with our estimate 
within the errors.

\subsection{$\alpha$-elements}

A number of lines for those elements formed through $\alpha$-capture, 
namely O, Mg, Si, Ca and Ti, were measured. For all these elements we note
a high level of homogeneity, with the star-to-star scatter consistent 
with the measured errors and without significant trends with $T_{eff}$.\\
The O analysis is based on the forbidden lines at $\lambda$6300.31 $\mathring{A}$ and
$\lambda$6363.79 $\mathring{A}$. 
These lines are not blended with 
telluric features, with the only exception of the line at $\lambda$6300.31 $\mathring{A}$ 
in the NGC 1978 spectra, which is blended with the telluric absorption line. 
For these stars this spurious contribution was removed by using the IRAF 
task TELLURIC and adopting as template spectrum an early type star.
At the UVES resolution, the $\lambda$6300.31 $\mathring{A}$ feature is well separated 
from the ScII line at $\lambda$6300.69 $\mathring{A}$ but contaminated by the very close
Ni transition at $\lambda$6300.34 $\mathring{A}$. In order to measure the correct 
oxygen abundance we used spectrum synthesis convolved with a Gaussian instrumental profile. 

To model the Ni line we used the measured abundance (see Sect. 4.1), while to model 
the various CN lines
we needed to assume C and N abundances ([C/Fe]=-0.5 dex and [N/Fe]=+0.5 dex) 
since not directly measurable.
However, it must be noted that the assumed C and N abundances in 
the typical range shown by RGB stars (e.g. 
-1$<$[C/Fe]$<$0.0 and 
0.0$<$[N/Fe]$<$+1) have only a marginal 
impact on the derived O abundance. 
For the other $\alpha$-elements we cross-checked the results derived by the EW 
measurements, by performing a synthetic spectrum fitting for some {\sl test} lines. 
This sanity check confirms the reliability of the derived abundances for these 
elements. 
Furthermore, the spectral region between $\lambda$6155 and $\lambda$6167 
$\mathring{A}$ (used to test the Ca abundances) includes the Ca line at 
$\lambda$6162.17 $\mathring{A}$ with strong damping wings, that are very sensitive to 
the electronic pressure and to the gravity but not sensitive to $T_{eff}$, 
$v_t$ and non-LTE effects \citep[see discussion in][]{mishen}. 
We are able to well-reproduce the wings shape of this line, confirming 
the reliability of the adopted gravities.

All 4 clusters show mildly subsolar [O/Fe] ratios (-0.04 --- -0.11 dex), 
with star-to-star scatter less than 0.10 dex. 
For NGC 1978  
\citet{hill00}  measured 
[O/Fe]=0.37 dex with a star-to-star scatter of 0.10 dex,
clearly in disagreement with our determination ([O/Fe]=-0.11 dex),
but this discrepancy can be again manly ascribed to the different 
iron content.\\
For the other elements, the [$\alpha$/Fe] turns out to be roughly solar, 
with a mild enhancement of [Mg/Fe] ($\sim$0.10-0.19 dex) and 
[Ti/Fe] in NGC~2173 (0.15 dex). \\

\subsection{s and r-process elements}

Several s-process elements, namely the light Y and Zr and the heavy 
Ba, La, Ce and Nd have been measured, together with Eu, a r-process element. 

The Ba abundance was derived by measuring the EWs of 
three lines. 
We tested the possible impact of the HFS by performing
a spectral synthesis using both the single component line and the separated HFS 
components taken from the linelist by \citet{pro00}. 
The inclusion of the HFS 
has a negligible ($\le$0.5\%) effect 
on the derived Ba abundance, 
due to the strong intensity of these lines, in which 
all the hyperfine components in the line core are completely saturated, 
reducing the HFS effects \citep[see the discussion in][]{mcw95}.
 This finding results in 
agreement with the analysis of the Ba line at $\lambda$6496.91 $\mathring{A}$ by 
\citet{norris97}.
This was also verified by mean of spectrum synthesis simulations of the 
three lines, using both the single component line and the separated HFS 
components taken from the linelist by \citet{pro00}. 
In Tab.~\ref{tab_ba} 
we report the comparison between the EWs measured on synthetic 
spectra of Ba lines computed both with and without HFS, for some target 
stars with different atmospheric parameters. HFS components are from 
\citet{pro00}. The very small differences confirm that the 
effect of inclusion of HFS is negligible for these lines.
Ce has an even atomic number {\it Z=58} and all of the isotopes have 
even neutron number {\it N}, hence there is no HFS. 
However,
an isotopic splitting 
is possible but we did not consider it since  
the calculations of \citet{aoki} show that it has a 
negligible impact on the derived abundance.  
The Nd line HFS can be neglected, because the only isotopes 
(the odd $^{143}Nd$ and $^{145}Nd$) 
that show it  account for $\sim$20\% \citep{hartog}
of the total abundance, only. 
The LaII line at $\lambda$ 6390.46 $\mathring{A}$ 
exhibits in all the target stars EWs small enough 
($\approx$70 m$\mathring{A}$) to safely ignore the HFS. 
Finally, Eu abundances have been derived by performing a 
spectrum synthesis fitting, 
by using the HFS parameters for the Eu at $\lambda$ 6645.11 
$\mathring{A}$ line by \citet{lawler}.
We noted as the average values of [Eu/Fe] in each cluster result 
very similar to the ones obtained by using the EW measurements 
(and ignoring 
possible HFS effects) but with a reduced star-to-star scatter, 
in agreement with the results obtained only from EWs by 
\citet{gratton06} and \citet{carretta07} for the Galactic GCs 
NGC 6441 and NGC 6338, respectively (whose metallicity is very similar 
to the mean metal abundance of the present sample, although based on 
higher S/N spectra).

In all the target clusters [Y/Fe] and [Zr/Fe] ratios result 
significantly depleted
($\le$-0.30 dex) with respect to the solar value. 
Heavy-s elements show enhanced 
($\sim$0.20-0.45 dex) 
[Ba/Fe], [La/Fe] and [Nd/Fe] ratios but 
[Ce/Fe] which turns out to be  
solar. Finally, all the clusters display an enhanced  
($>$0.30 dex) [Eu/Fe] abundance ratio.

\section{Discussion}
		
All the 4 LMC globulars analysed here, 
belonging to the same age population but located 
in different regions of the LMC disk, result to be {\sl metal-rich},
with a mean metallicity of [Fe/H]=-0.38 dex
(rms=0.09 dex). This finding confirms the previous low-resolution 
analysis based on the 
Ca II triplet by \citet{ols91} and \citet{aaron}, that showed as the 
young and intermediate-age LMC clusters exhibit a very narrow metallicity 
distribution 
\citep[][estimated a mean metallicity of -0.48 dex with a rms=0.09 dex from 23 intermediate-age clusters]{aaron}.
Our metallicities are also consistent with the mean metallicity of the 
LMC Bar ([Fe/H]=-0.37 dex) by \citet{cole05} 
and of the LMC disk ([Fe/H]$\sim$-0.5 dex) by \citet{carrera07}.
The theoretical 
scenario for the formation of the LMC and its GCs drawn by \citet{bekki} 
indicates as the efficient GC formation does not occur until the LMC and 
the SMC start to interact violently and closely ($\sim$3 Gyr ago). Moreover, 
also the formation of the Bar is predicted to occur in the last $\sim$5 Gyr.
The similar iron content between the LMC GC system and the LMC Bar 
seems to confirm this hypothesis. 
		
Our chemical analysis also evidences a high degree of homogeneity for all 
the elements. 
Even the abundances of Na and Al 
show a low dispersion, at variance 
with Galactic GCs which show strong 
O-Na and Mg-Al anticorrelations 
\citep[see e.g.][ for an extensive review]{gsc}. 

The depletion (by a factor of 2-3) of [Na/Fe] and [Al/Fe]  
abundance ratios with respect to the solar and Galactic thin disk values, 
is consistent with the one observed in the LMC and Sgr fields.
These two elements are likely connected to the SNII, because 
their main production sites are C and Ne burning
\citep[see][]{pagel, matteucci03}, respectively. 
Another possible channel to produce Na and Al are the p-captures in the 
intermediate-mass AGB stars (NeNa and MgAl cycles).  
However,
the high degree of homogeneity of their abundance in the LMC clusters 
and the lack of clear Na-O and Mg-Al anti-correlations
seem to favor the SNII channel for their production.
Also, since the Na and Al yields depend on the 
neutron excess and increase with metallicity
\citep{pagel},  
under-abundant [Al/Fe] and [Na/Fe] ratios suggest 
that the gas from which the LMC clusters formed, 
should have been enriched by relatively low-metallicity SNII.

Also the $\alpha$-elements are produced mainly by high-mass stars 
which end their short life exploding as SN II, but 
at variance with Na and Al, their production factors are poorly sensitive 
to metallicity. 
The [$\alpha$/Fe] ratio represents a powerful diagnostics
to clarify the relative role played 
by SNII (producers of $\alpha$-elements) and SNIa (main producers of Fe) in 
the chemical enrichment process.
Indeed, there is time delay \citep{tinsley} between the explosion of SNII, 
occurring since the onset of the 
star formation event, and SNIa, which happen later on \citep{greg}.
The roughly solar [$\alpha$/Fe] abundance ratios measured in the 
LMC GCs well match those found in the LMC field and MW 
thin disk intermediate-age populations and
are consistent with a standard scenario,  
where SNIa had enough time to significantly enrich the gas with iron.
Some depletion of [Mg/Fe], 
[Ca/Fe] and [Ti/Fe] is observed in the Sgr stars.

The bulk of the iron-peak elements are produced 
by the SNIa, from stars with intermediate-mass progenitors and located in 
single-degenerate binary systems \citep{iwamoto}, 
or from double-degenerate binary systems \citep{it84}.  
Our derived abundances for these elements well trace iron, 
as do the LMC field and MW thin disk stars. 
We only note a mild discrepancy between our 
[Ni/Fe] solar ratio and the  
slightly underabundant ([Ni/Fe]$\sim$-0.2 dex) values by \citet{pompeia06}, also 
observed in the Sgr stars that show a significant depletion (by a factor of 2-3) 
of iron-peak elements.
It is interesting to note that the old LMC clusters analysed by 
\citet{johnson06} show a general depletion of the [iron-peak/Fe] ratios
(in particular [V/Fe] and [Ni/Fe]):
actually, an explanation for such a depletion for the iron-peak 
elements is still lacking.

The elements heavier than the iron-peak group are not built up from 
thermonuclear burning but
via a sequences of neutron captures on seed Fe or Ni nuclei. 
If the time-scale of the neutron
capture sequence is longer than the typical time-scale of the $\beta$-decay,
 the resulting
elements are called {\sl slow} or s-process elements, while in case 
of fast neutron capture, 
the elements are called {\sl rapid} or r-process elements.
The s-process elements are mainly produced by low-mass ($\sim$1-4 $M_{\odot}$) 
AGB stars during the thermal instabilities developing above the 
quiescent He-burning shell (the so-called {\sl main}-component), 
with a minor contribution by the high mass stars 
(the so-called {\sl weak}-component) \citep[see][]{busso,travaglio}.  
The bulk of these neutron captures are 
connected to the  
$^{13}C(\alpha,n)^{16}O$  and 
$^{22}Ne(\alpha,n)^{25}Mg$ reactions, which are major sources of neutrons.

The behaviour of the s-process elements in the LMC clusters 
appears to be {\sl dichotomic}, with a deficiency 
of light s-elements (Y and Zr) and an enhancement
of heavy ones (Ba, La and Nd),  with 
the only exception of Ce, that shows a solar [Ce/Fe] abundance ratio.
The [Ba/Y] abundance ratio represents a powerful diagnostic of the relative 
contribution of the heavy to the light s-process elements \citep[see][]{venn}. 
In our LMC clusters [Ba/Y] is enhanced by $\sim$0.9-1 dex (see Fig. 8): 
similar values have been observed also in the LMC field \citep{hill95, pompeia06} and
in Sgr \citep{sbordone}, 
but not in the MW, where the [Ba/Y] ratio is solar at most.
The interpretation of these abundance patterns is complicated by the 
complexity (and uncertainty) of the involved nucleosynthesis.
Theoretical models \citep{busso,travaglio} indicate that the AGB yields 
could be metallicity-dependent. In particular, the heavy-s elements 
have their maximum production factor at lower metallicities than the  
light-s ones. 
Hence, a high [Ba/Y] ratio could suggest a major pollution of the gas
by low-metallicity AGB stars. 
 Moreover, abundance patterns for [Y/Fe] and [Ba/Fe] consistent 
with the Galactic values have been observed in the old LMC cluster 
by \citet{johnson06}. Being these objects the first ones 
formed in the LMC, these clusters have been not contaminated 
by the AGB stars, because  the low-mass AGB stars had no time 
to evolve and incorporate completely their yields in the 
interstellar medium (differently to the intermediate-age 
clusters).

The [Eu/Fe] abundance ratio (see Fig.~8) measured in the LMC clusters and Sgr stars 
is enhanced by a factor of two  with respect to the Galactic thin disk value. 
This is somewhat puzzling and inconsistent with the solar $<[\alpha/Fe]>$ measured 
in all the three environments.
Indeed, Eu is a typical r-process element \citep{arl, burris}, whose 
most promising site of nucleosynthesis are SNII
(SNII with low \citep[M$<$11 $M_{\odot}$,][]{cowan} mass progenitors are  
interesting candidates), although 
other alternative sites are possible \citep[see e.g.][]{cowan}.
Such an anomalous high [Eu/Fe] abundance ratio seems to suggest that 
in the LMC clusters and Sgr the Eu
is not or not only synthesized in a similar fashion as the $\alpha$-elements.

Finally, the ratio between the s-process elements (which are predominantly formed 
through slow neutron captures, with a minor contribution from rapid neutron captures) 
and Eu (a pure r-process element) 
represents a powerful diagnostics
to estimate the relative contribution of the different neutron-capture 
processes.
The theoretical solar [Ba/Eu] in case of pure rapid neutron captures 
turns out to be -0.70 dex \citep{arl}.  
The [Ba/Eu]$\approx$0.0 dex abundance ratios measured in the 
LMC clusters as well as in the Galactic thin disk  
and Sgr stars, suggest 
that s-process elements should be mainly produced by AGB stars through 
slow neutron captures, with a minor (if any) 
contribution from massive SNII through rapid neutron captures.

At variance, 
the strong [Y/Eu] ($\le$-0.7 dex) depletion measured 
in the LMC clusters and Sgr stars
is very different from the higher values 
([Y/Eu]$\sim$-0.20 dex 
at [Fe/H]=-0.40 dex, see Fig. 8)
observed in the thin disk stars.
This may suggest that rapid neutron captures 
can also have a different role in 
the production of light and heavy s-process elements, 
(as suggested by \citet{venn} in order to explain similar patterns 
in the dSphs) but this seems to be environment-dependent.
Alternatively, galactic winds could have been more effective
in removing these elements from the LMC and Sgr  
\citep[see e.g.][]{matt83}.

\section{Conclusions}

We performed a detailed abundance analysis of the most 
important chemical elements of 
27 giant stars members of 4 LMC intermediate-age clusters.
We compared the inferred abundance patterns with those 
of other intermediate age populations in different 
galactic environments, namely the LMC field, the Galactic 
thin disk and the Sgr dwarf spheroidal. 
Such an analysis allows us to 
obtain important information about the chemical 
properties of the intermediate-age population of the LMC:

(1) As unequivocally traced by both field and cluster stars,
the intermediate-age population of the LMC is metal-rich, 
with an average iron content between 1/3 and half solar. 

(2) The interstellar medium from which these stars 
formed had the time to be significantly 
enriched by SNIa and AGB star ejecta, as traced by the 
[$\alpha$/Fe] and [s-process/Fe] abundance patterns.

(3) An enhanced pollution of the gas (from which 
these clusters formed) 
by SNII and AGB stars with low metallicity 
could explain either the depletion of [Al/Fe] and 
[Na/Fe] and the enhancement of the [Ba/Y] abundance ratios
with respect to the values measured in 
the Galactic thin disk stars. 

(4) The lack of clear O-Na and Mg-Al anti-correlations seems 
to indicate that the 
studied LMC clusters did not undergo 
appreciable self-enrichment, as likely did the old Galactic GCs, 
but this has to be proven on a better 
statistical ground.

(5) The enhanced [Eu/Fe] ratios appear to be in contradiction 
with the solar [$\alpha$/Fe] ratio, despite the same nucleosinthetic 
site (massive stars). This decoupling between r- and $\alpha$-elements 
seems to be a distinctive feature of several extragalactic 
environments (LMC, SMC, Sgr, dSphs).

The chemical analysis of these clusters
provides an overall picture of the metal-rich, intermediate-age 
component of the LMC cluster system, remarkably
different with respect to the Galactic field populations of similar 
ages and metallicities. We found similar [$\alpha$/Fe] and 
[iron-peak/Fe] ratios and discrepant light Z-odd and 
neutron-capture elements. 
Moreover, the comparison with 
the Sgr dSph evidences some likeness ([Na/Fe], [Al/Fe] and the 
neutron capture elements pattern) but also several differences 
(as the average value of the [$\alpha$/Fe] and
[iron-peak/Fe] ratios).  
Our results point toward a scenario of 
chemical evolution dominated by 
previous generations of low-metallicity stars, able to produce 
the observed depletion of light Z-odd elements as well as the behaviour of 
the s-process elements.
The extension of our study to additional younger and older 
LMC clusters will provide new insight towards 
the understanding of the LMC formation and 
chemical enrichment history.

\acknowledgements  

This research was supported by the Ministero dell'Istruzione dell'Universit\'a 
e della Ricerca and it is part of the {\it "Progetti Strategici d'Ateneo 2006"} 
granted by the Bologna University. We warmly thank the anonymous referee for his/her suggestions.
Moreover, we warmly thank Elena Pancino, Angela Bragaglia and Raffaele Gratton for 
useful suggestions and discussions, and Elena Valenti and Elena Sabbi 
for their collaboration in the preparation of the observations.

\begin{figure}
\plotone{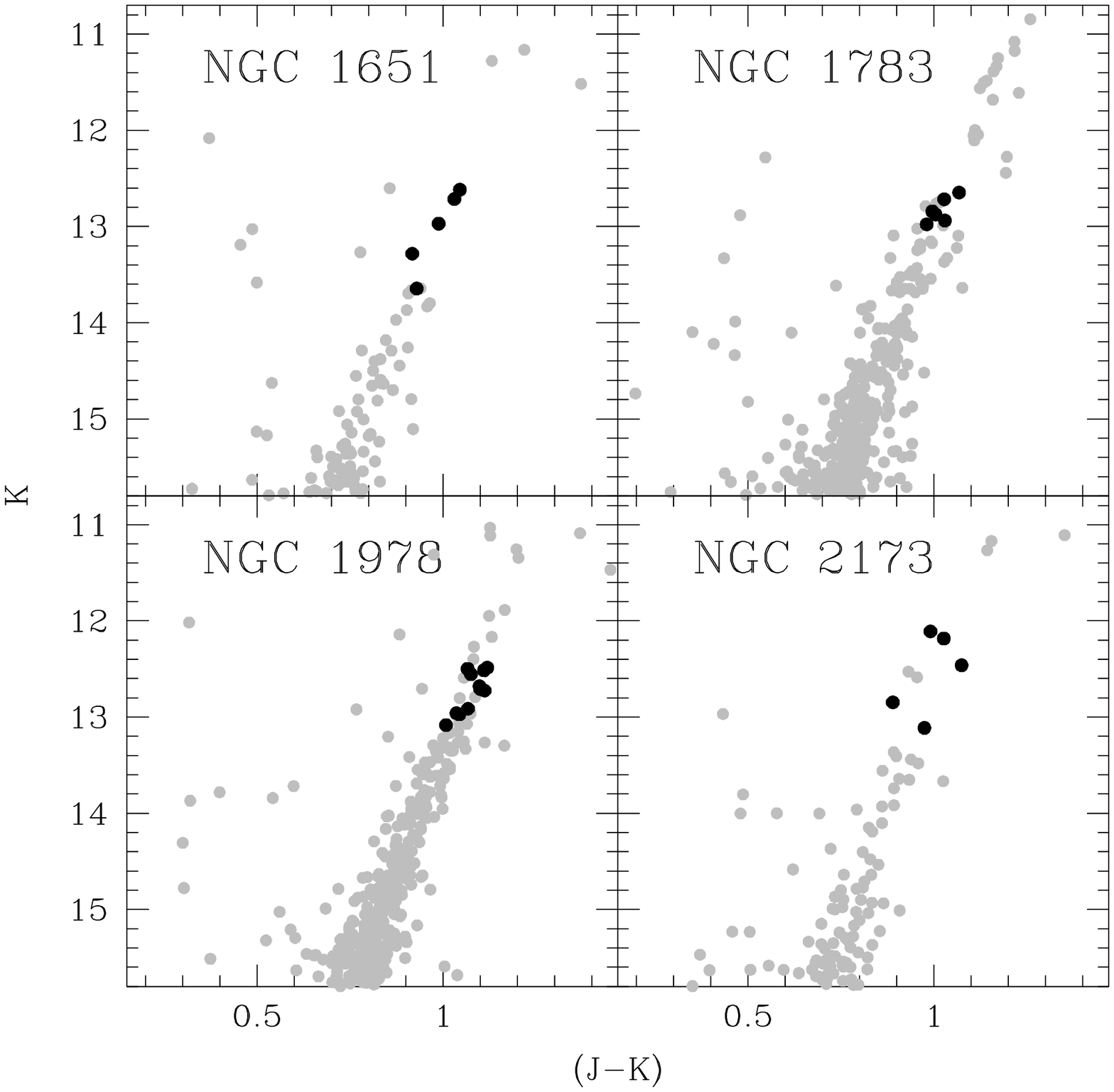} 
\caption{K, J-K color-magnitude diagrams for the upper RGB of the 4 observed LMC clusters 
\citep{m06a}: the black 
points indicate the target stars of the present work.}
\label{f}
\end{figure}

\begin{figure}
\plotone{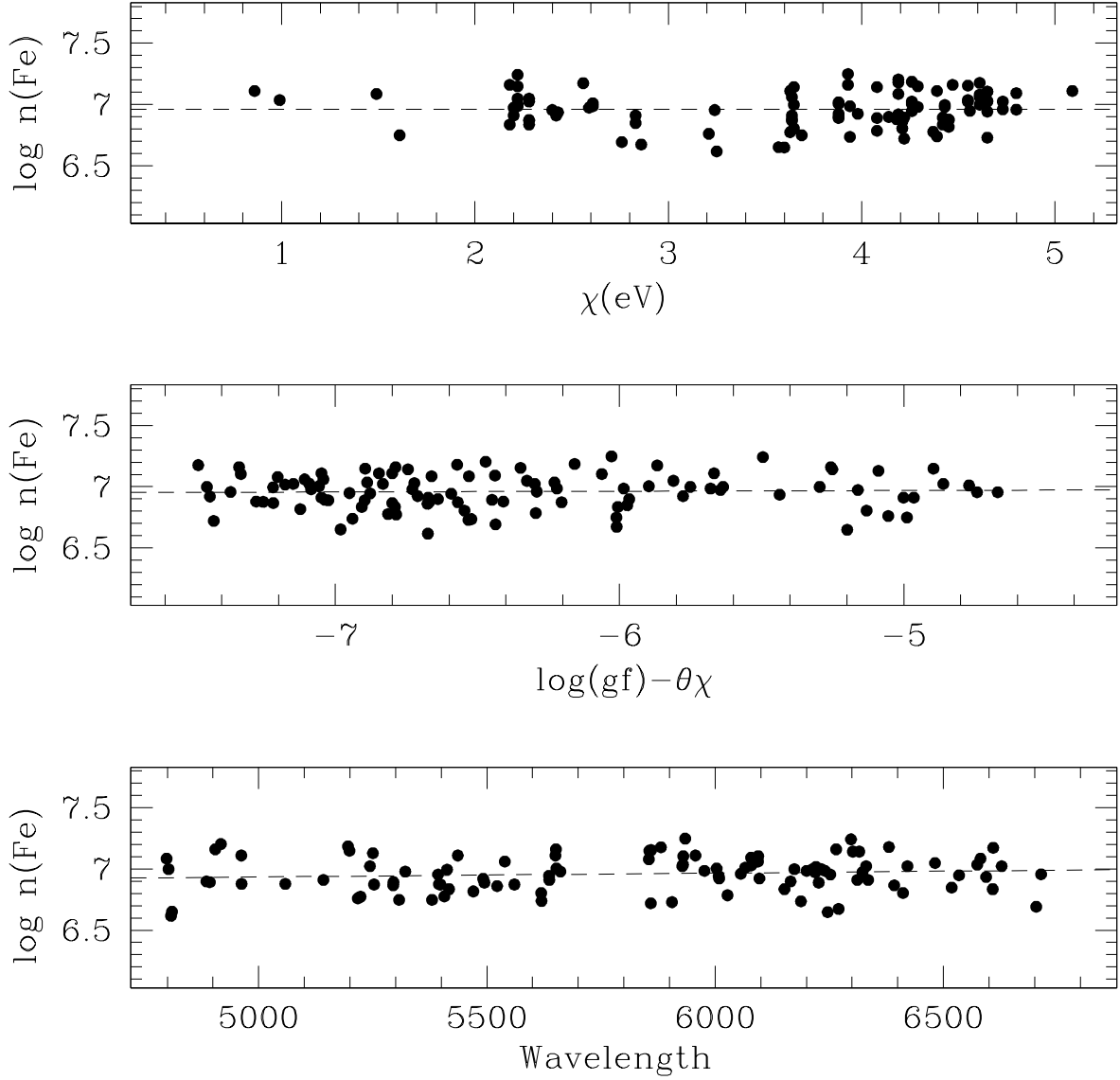} 
\caption{The trend of the derived abundances from Fe I lines 
for the star NGC 2173-8 as a function of the excitation potential 
$\chi$ (upper panel), the expected  
line strength (middle panel) and the wavelength (lower panel). 
The dashed lines represent the linear fit to each distribution.}
\label{f}
\end{figure}

\begin{figure}
\plotone{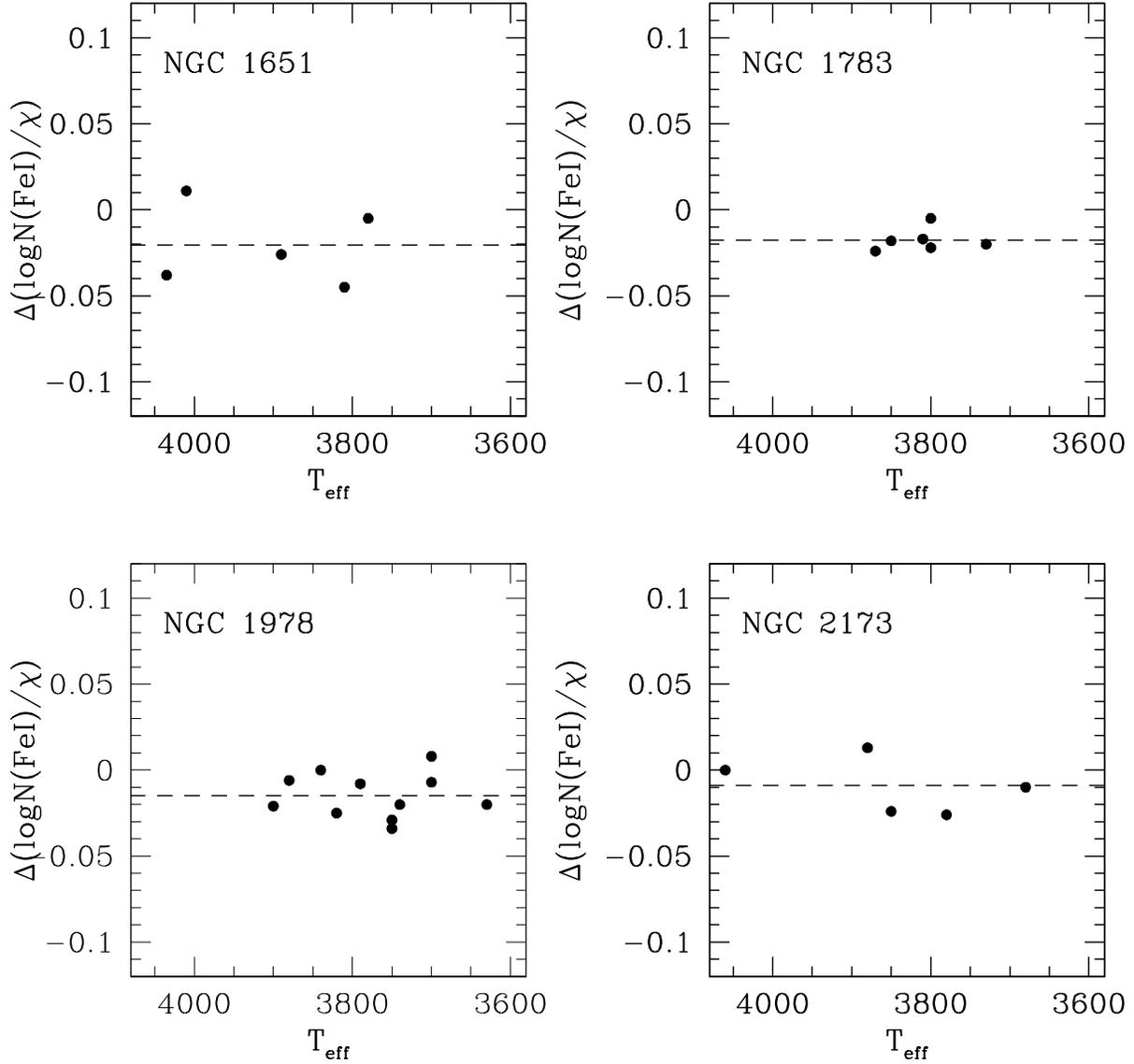} 
\caption{The slopes of the relationship between the neutral iron abundances and the excitation potential 
$\chi$ for individual stars in each cluster as a function of $T_{eff}$.
Dashed lines represent the average slope for each cluster.}
\label{f}
\end{figure}

\begin{figure}
\plotone{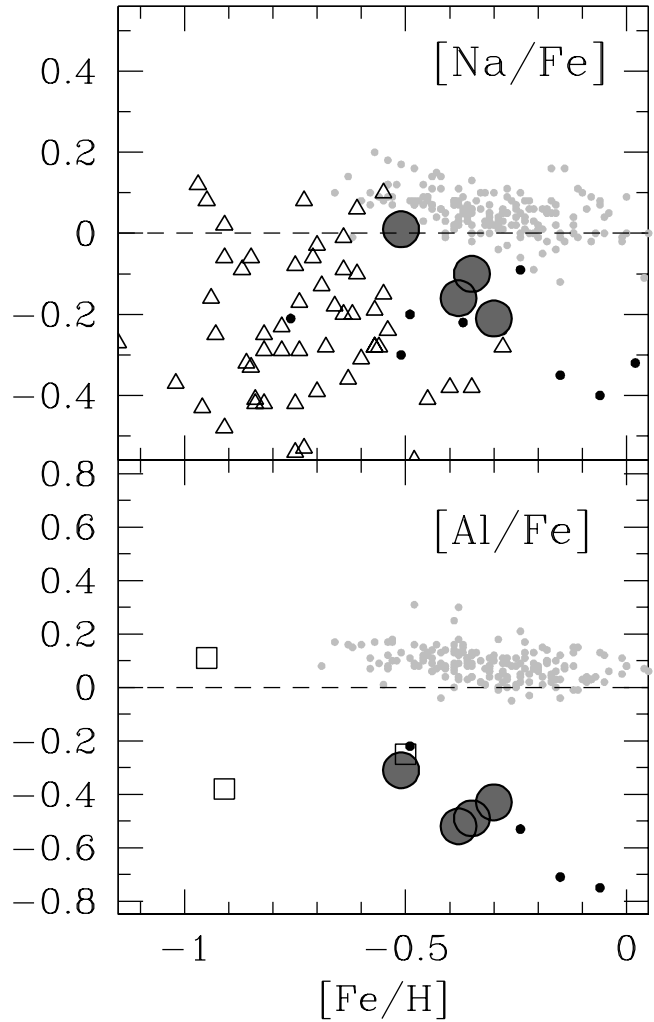} 
\caption{The trend of [Na/Fe] and [Al/Fe] as a function of [Fe/H] (upper and lower panel
respectively) for the 4 analysed LMC clusters (big grey points). For comparison, 
previous determinations of these abundance ratios in the LMC field (empty triangles from \citet{pompeia06}), 
other LMC clusters (empty squares from \citet{hill00}), 
the Galactic thin disk (little grey points from \citet{reddy}) and Sgr (little black points 
from \citet{boni00}, \citet{monaco05}, \citet{monaco07} and \citet{sbordone}) are also plotted.}
\label{f}
\end{figure}

\begin{figure}
\plotone{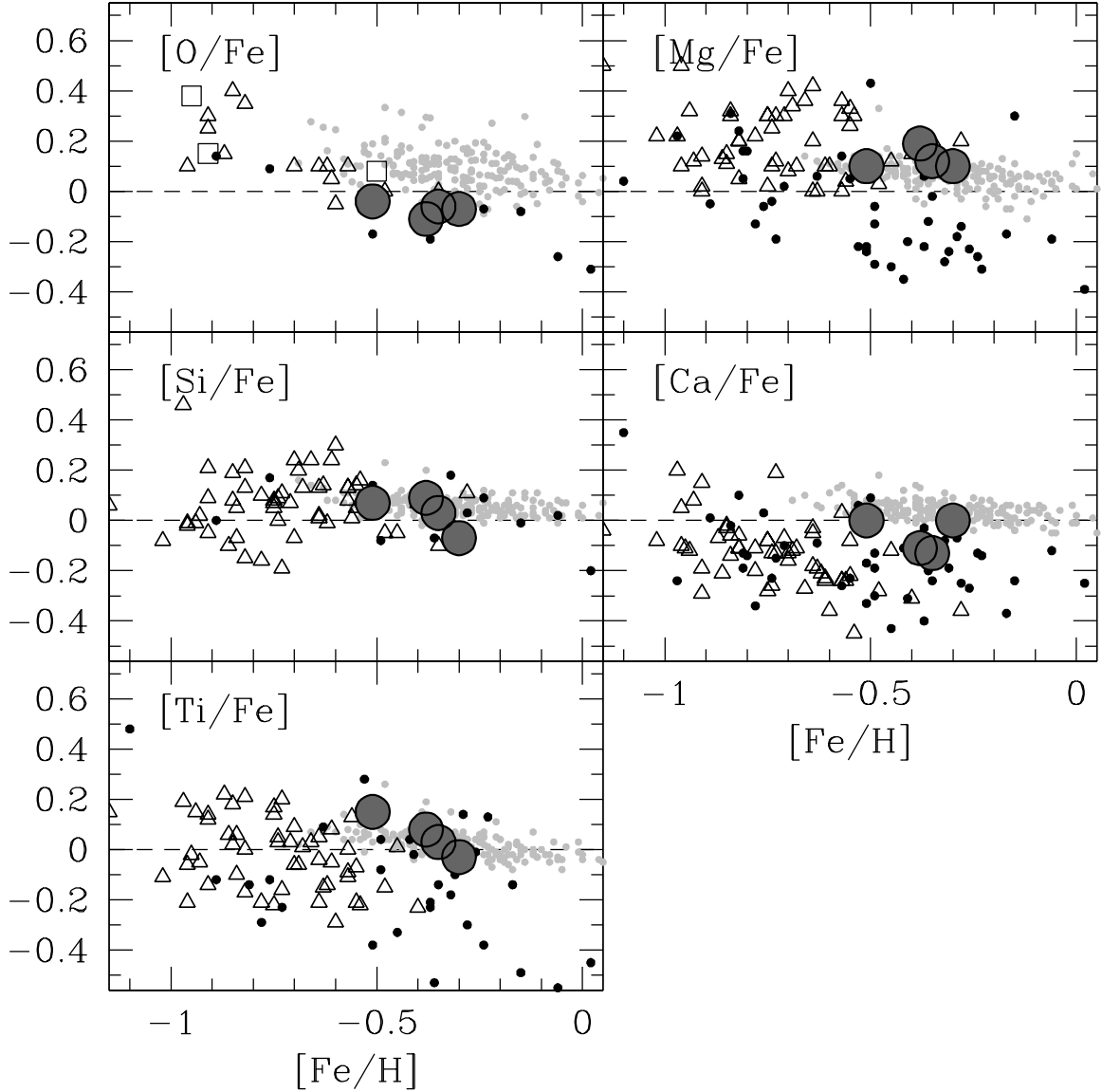} 
\caption{The trend of $\alpha$-elements ([O, Mg, Si, Ca, Ti/Fe]) 
as a function of [Fe/H] for the 4 analysed LMC clusters (same symbols and references of Fig.
4).}
\label{f}
\end{figure}

\begin{figure}
\plotone{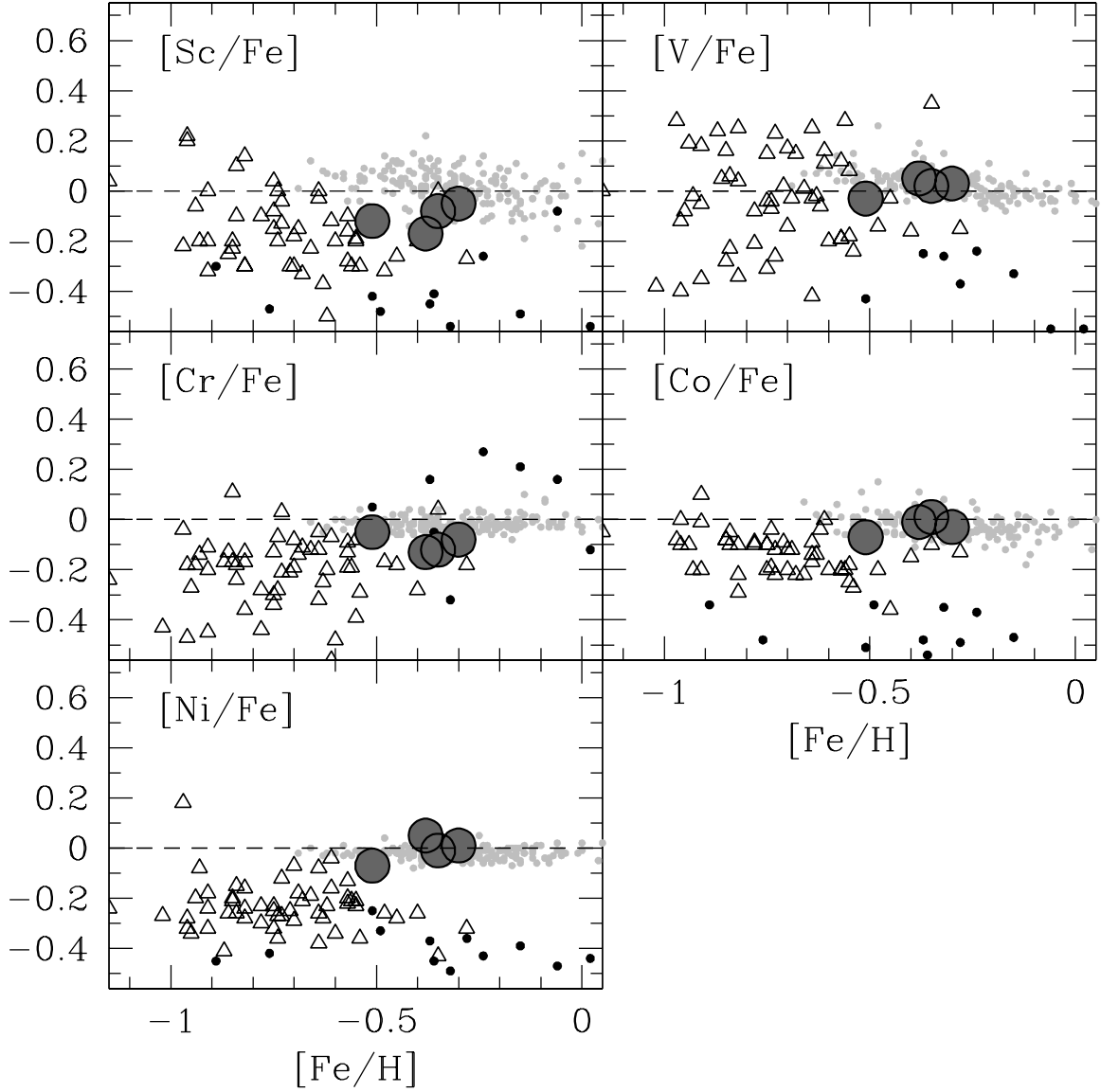} 
\caption{The trend of iron-peak elements ([Sc, V, Cr, Co, Ni/Fe]) as a function of [Fe/H] (same symbols 
and references of Fig. 4).}
\label{f}
\end{figure}

\begin{figure}
\plotone{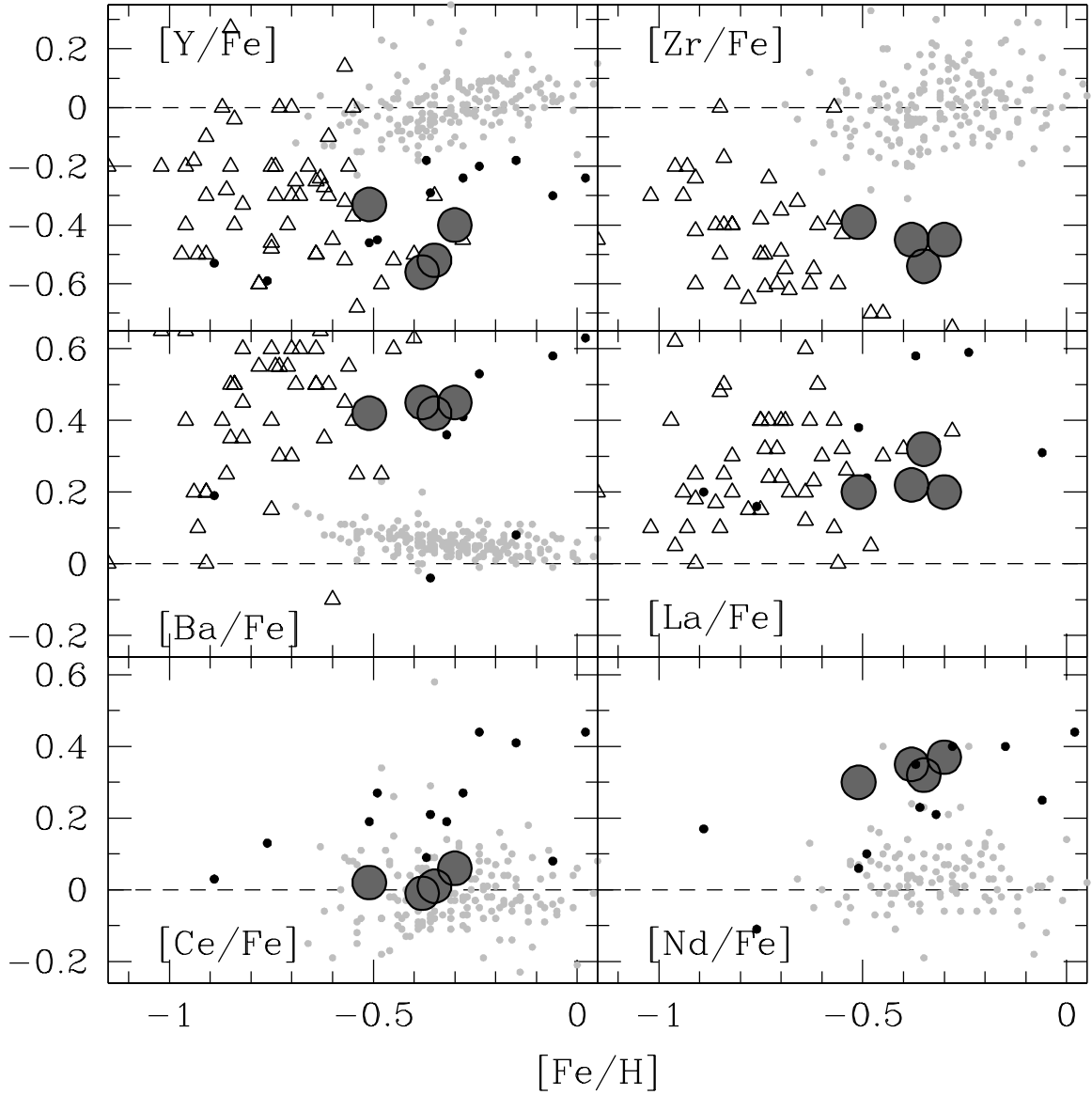} 
\caption{The trend of light ([Y,Zr/Fe]) and heavy ([Ba, La, Ce, Nd/Fe]) s-process elements 
as a function of [Fe/H] (same symbols and references of Fig. 4).}
\label{f}
\end{figure}

\begin{figure}
\plotone{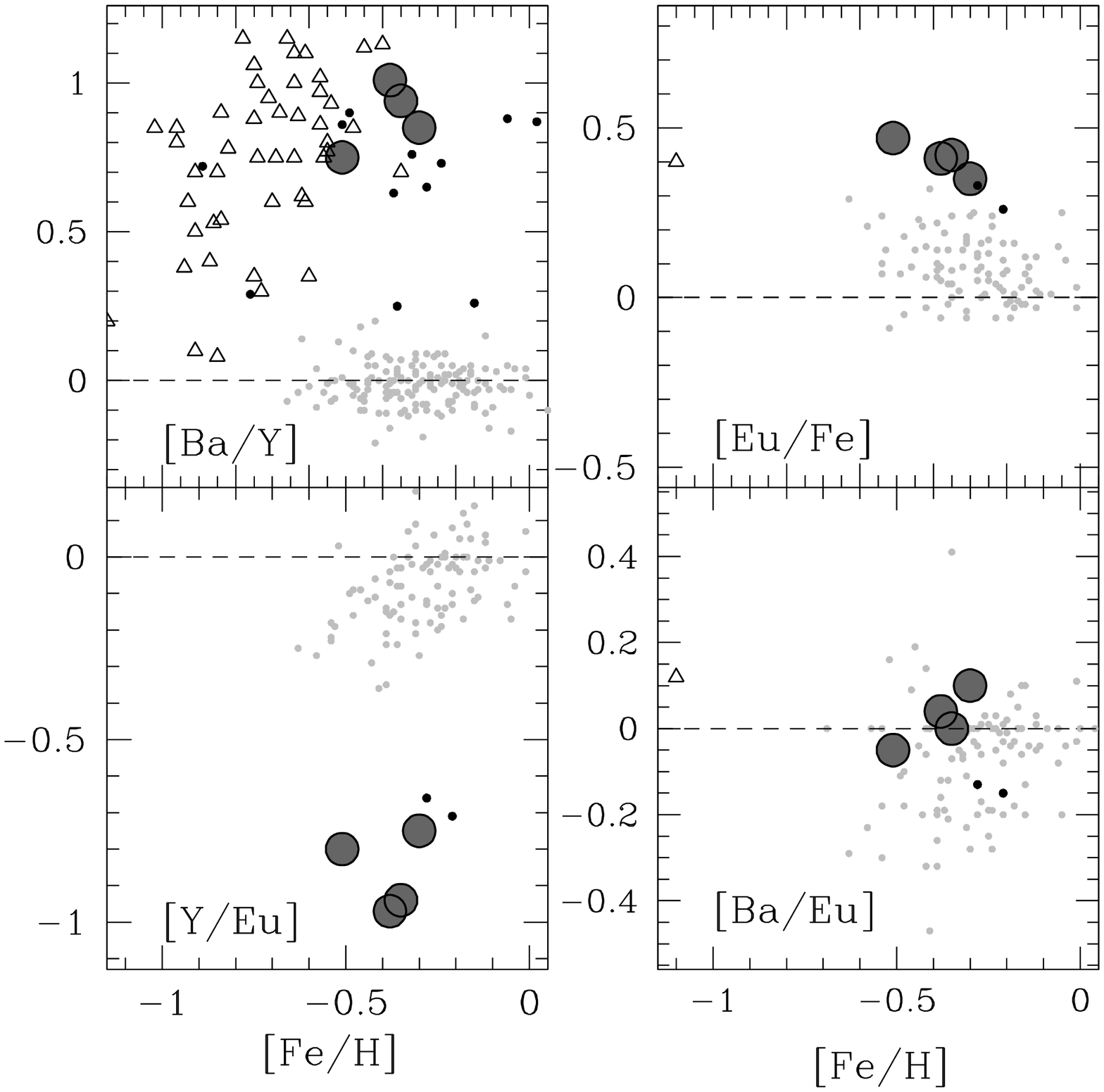} 
\caption{The trend of [Ba/Y], [Eu/Fe], [Y/Eu] and [Ba/Eu] as a function of [Fe/H]
(same symbols and references of Fig. 4).}
\label{f}
\end{figure}

\begin{deluxetable}{lcccccc} 
\footnotesize
\tablecolumns{7} 
\tablewidth{0pc} 
\tablecaption{Informations about the target stars.} 
\tablehead{ 
\colhead{Star ID}  & \colhead{$S/N$}& \colhead{$V_{helio}$}
 &  \colhead{$J_0$} & \colhead{$(J-K)_0$} & \colhead{RA(J2000)} & \colhead{Dec(J2000)}\\
& & \colhead{(km/s)}& &  & &  }
\startdata 
   1651-6 &   35     &	234.3  & 13.58  &  0.99&  69.3880040   &   -70.6012635    \\
   1651-8 &   30     &  227.3  & 13.66  &  0.98&  69.3799588   &   -70.5734344 \\
   1651-10 &   30    & 	232.1  & 13.87  &  0.94&  69.3844841   &   -70.5838366 \\
   1651-12 &   30    &  235.7  & 14.11  &  0.86&  69.3573261   &   -70.5738910     \\
   1651-16 &   25    &  236.2  & 14.49  &  0.88&  69.3824681   &   -70.5959887   \\
\hline
   1783-22 &   35    &  277.4  & 13.63   &  1.02&  74.7264895	&   -65.9723560    \\
   1783-23 &   30    &  275.1  & 13.66   &  0.98&  74.7793127	&   -65.9862323 \\  
   1783-29 &   30    &  275.2  & 13.75   &  0.94&  74.7830715	&   -65.9957701 \\
   1783-30 &   30    &  281.2  & 13.79   &  0.95&  74.8010628	&   -65.9629302    \\
   1783-32 &   30    &  277.9  & 13.87   &  0.93&  74.7707871	&   -65.9799639 \\   
   1783-33 &   35    &  278.8  & 13.88   &  0.98&  74.8011665	&   -65.9906700   \\
\hline  
   1978-21 &   35    &  295.5   & 13.52   &  1.07&  82.1515305   &   -66.2322134      \\
   1978-22 &   40    &  290.6   & 13.48   &  1.01&  82.2014424   &   -66.2339097   \\
   1978-23 &   35    &  288.7   & 13.54   &  1.06&  82.2092677   &   -66.2568186  \\
   1978-24 &   30    &  291.5   & 13.54   &  1.02&  82.1915173   &   -66.2387280 \\  
   1978-26 &   30    &  296.3   & 13.69   &  1.05&  82.1751055   &   -66.2325960   \\
   1978-28 &   35    &  292.3   & 13.72   &  1.05&  82.1774043   &   -66.2079169	 \\
   1978-29 &   35    &  298.4   & 13.75   &  1.06&  82.1906198   &   -66.2420488 \\
   1978-32 &   30    &  290.5   & 13.89   &  1.01&  82.1764751   &   -66.2351731  \\
   1978-34 &   20    &  292.1   & 13.91   &  0.98&  82.2041703   &   -66.2277628  \\
   1978-38 &   30    &  297.1   & 13.93   &  0.99&  82.2221112   &   -66.2352105 \\
   1978-42 &   35    &  291.5   & 14.00   &  0.96&  82.1706985   &   -66.2461504     \\
\hline
    2173-4 &   50    & 237.6	& 12.95    &  0.86 &    89.4861621 &  -72.9749781  \\
    2173-5 &   50    & 234.8	& 13.15    &  0.99 &    89.4910121 &  -72.9652585     \\
    2173-6 &   40    & 237.3	& 13.47    &  1.04 &    89.5475844 &  -72.9757905   \\
    2173-8 &   55    & 237.9    & 13.68    &  0.85 &    89.4955156 &  -72.9785015   \\
   2173-10 &   40    & 236.3	& 14.03    &  0.94 &    89.4816674 &  -72.9815157    \\

\enddata
\tablecomments{The adopted E(B-V) reddening values \citep{pers83} are 0.1 for NGC~1651, NGC~1783 and 
NGC~1978, and 0.07 for NGC~2173.}
\end{deluxetable}

\begin{deluxetable}{rrrrrrrrrrr} 
\tablecolumns{10} 
\tablewidth{0pc}  
\tablecaption{Equivalent widths from UVES spectra for the sample stars (in electronic form).}
\tablehead{ 
\colhead{El.}  & \colhead{$\lambda$}& 
\colhead{E.P.}
 &  \colhead{log gf} & 
\colhead{1651-6}  & \colhead{1651-8} & \colhead{1651-10}  & \colhead{1651-12} 
& \colhead{1651-16} & \colhead{1783-22} \\ 
 & \colhead{($\mathring{A}$)} & \colhead{(eV)} &  & \colhead{(m$\mathring{A}$)} 
 & \colhead{(m$\mathring{A}$)} &  \colhead{(m$\mathring{A}$)}  & \colhead{(m$\mathring{A}$)} 
 & \colhead{(m$\mathring{A}$)} & \colhead{(m$\mathring{A}$)}  
}
\startdata 
 \hline
    O I &  6300.31    & 0.00 &  -9.75    &   syn       &       syn      &	syn   &    syn  &     syn      &     syn    	           \\
   O I &  6363.79    & 0.02 & -10.25    &   syn       &       syn      &	syn   &    syn  &     syn      &     syn    	          \\
   Na I & 5682.65    & 2.10 &  -0.67    &   0.0       &     102.7      &      129.9   &    146.0&  119.7       &    119.1   	       \\
   Na I & 5688.22    & 2.10 &  -0.37    & 142.0       &     130.7      &      144.8   &    159.8&  145.1       &    142.1   	       \\
   Na I & 6154.23    & 2.10 &  -1.57    &  74.1       &      41.0      &       69.7   &     74.5&   45.2       &     80.4   	       \\
   Na I & 6160.75    & 2.10 &  -1.26    &  88.0       &      73.5      &       88.7   &     87.7& 88.6         &    102.1   	      \\
   Mg I & 5528.42    & 4.34 &  -0.52    & 257.2       &     247.7      &      252.2   &   270.5 & 254.1	       &    227.6   	         \\
   Mg I & 5528.42    & 4.34 &  -0.52    &   0.0       &	     0.0       &       0.0    &   0.0	&  0.0         &   131.9        \\
   Mg I & 6318.71    & 5.11 &  -1.94    &  69.1       &      55.9      &       63.5   &     65.3&   77.3       &    0.0 	         \\
   Mg I & 6319.24    & 5.11 &  -2.16    &  44.0       &       0.0      &        0.0   &     46.4&    43.8      &    43.7	         \\
\enddata 												    
\end{deluxetable}

\begin{deluxetable}{lcccccc} 
\footnotesize
\tablecolumns{7} 
\tablewidth{0pc}  
\tablecaption{Sensitivities of the abundance ratios to the variation of the 
atmospheric parameters ($T_{eff}$, log~g, [A/H], $v_t$), with 
the corresponding average number of used lines ($<N>$) and the error 
of the abundance associated to the EW ($\sigma_{EW}$), 
as computed for star NGC~1783-29.}
\tablehead{ 
\colhead{Ratio} & \colhead{$T_{eff}$} 
& \colhead{log{g}}
 &  \colhead{[A/H]} & 
\colhead{$v_t$} & \colhead{$<N>$} & \colhead{$\sigma_{EW}$}  \\
&  \colhead{(+100 K)} & \colhead{(+0.2 dex)} & \colhead{(+0.1 dex)} &\colhead{(+0.2 km/s)} & &
}
\startdata 
  $[O/Fe]$     &    0.028      &     0.042    &     0.040      &   -0.016    &  2    &  0.127	    \\
  $[Na/Fe]$    &    0.085      &    -0.010    &     0.006      &   -0.075    &  4    &  0.090	    \\
  $[Mg/Fe]$    &    0.004      &     0.001    &     0.027      &   -0.059    &  4    &  0.090	    \\
  $[Al/Fe]$    &    0.069      &     0.003    &     0.005      &   -0.047    &  2    &  0.127	    \\
  $[Si/Fe]$    &   -0.113      &     0.027    &     0.025      &   -0.047    &  4    &  0.090	    \\
  $[Ca/Fe]$    &    0.063      &    -0.060    &    -0.032      &   -0.185    & 13    &  0.050	    \\
  $[Sc/Fe]$II  &    0.018      &     0.101    &     0.093      &   -0.091    &  5    &  0.080	   \\
  $[Ti/Fe]$    &    0.168      &     0.016    &     0.031      &   -0.167    & 13    &  0.050	    \\
  $[V/Fe]$     &    0.156      &     0.004    &     0.024      &   -0.195    &  7    &  0.068	   \\
  $[Cr/Fe]$    &    0.138      &     0.047    &     0.058      &   -0.073    & 14    &  0.048	    \\
  $[Fe/H]$I    &   -0.016      &     0.019    &     0.028      &   -0.114    & 86    &  0.019	   \\
  $[Fe/H]$II   &   -0.210      &     0.068    &     0.043      &   -0.063    &  2    &  0.127	   \\
  $[Co/Fe]$    &   -0.033      &     0.024    &     0.026      &   -0.117    &  3    &  0.104	   \\
  $[Ni/Fe]$    &   -0.040      &     0.028    &     0.028      &   -0.099    & 25    &  0.036	    \\
  $[Y/Fe]$II   &   -0.016      &     0.081    &     0.034      &   -0.160    &  3    &  0.104	  \\
  $[Zr/Fe]$    &    0.160      &     0.038    &     0.021      &   -0.154    &  3    &  0.104	  \\
  $[Ba/Fe]$II  &    0.019      &     0.056    &     0.050      &   -0.090    &  3    &  0.104	  \\
  $[La/H]$II   &    0.018      &     0.087    &     0.037      &   -0.113    &  1    &  0.180	  \\
  $[Ce/H]$II   &    0.013      &     0.082    &     0.038      &   -0.062    &  1    &  0.180	  \\
  $[Nd/Fe]$II  &    0.019      &     0.079    &     0.038      &   -0.166    &  2    &  0.127	  \\
  $[Eu/Fe]$II  &   -0.014      &     0.083    &     0.036      &   -0.053    &  1    &  0.180	  \\
\enddata 
\end{deluxetable}

\begin{deluxetable}{ccccccccccc}
\footnotesize
\tablecolumns{11} 
\tablewidth{0pc}  
\tablecaption{Adopted atmospheric parameters and inferred neutral and ionized iron abundances. 
Adopted reference solar are log~n(Fe I)=7.54 and log~n(Fe II)=7.49.}
\tablehead{ 
\colhead{Star ID}  & \colhead{$T_{eff}$}& \colhead{$log{g}$}
 &  \colhead{[A/H]} & 
\colhead{$v_t$} & \colhead{n}  & \colhead{[Fe/H]I} & \colhead{rms} & \colhead{n} & \colhead{[Fe/H]II} & \colhead{rms} \\
&  \colhead{(K)} & \colhead{(dex)} &  \colhead{(dex)}&\colhead{(km/s)}  &  &(dex) &(dex) &  & (dex) &(dex)
}
\startdata 
     1651-6 &   3780  & 0.76 & -0.25   &  1.48    & 79   &  -0.27  & 0.19   &---& ---	&  ---      \\
     1651-8 &   3810  & 0.81 & -0.40   &  1.52    & 81   &  -0.41  & 0.20   & 2	& -0.18 & 0.01    \\
    1651-10 &   3890  & 0.92 & -0.32   &  1.55    & 69   &  -0.32  & 0.19   & 2	& -0.22 & 0.16  \\
    1651-12 &   4035  & 1.07 & -0.31   &  1.70    & 89   &  -0.31  & 0.19   & 2	& -0.20 & 0.10  \\
    1651-16 &   4010  & 1.20 & -0.20   &  1.49    & 74   &  -0.21  & 0.15   & 2	& -0.15 & 0.15  \\
\hline
    1783-22 &   3730  & 0.80 & -0.27   &  1.35    & 79   &  -0.26  & 0.19   & 3	& -0.35 & 0.05 \\
    1783-23 &   3810  & 0.85 & -0.35   &  1.38    & 86   &  -0.34  & 0.17   &---& ---	& ---	    \\
    1783-29 &   3870  & 0.91 & -0.37   &  1.32    & 101  &  -0.38  & 0.20   & 2 & -0.25 & 0.13  \\
    1783-30 &   3850  & 0.92 & -0.36   &  1.37    & 79   &  -0.36  & 0.20   & 1	& -0.27 &  ---     \\
    1783-32 &   3800  & 0.92 & -0.30   &  1.22    & 95   &  -0.31  & 0.19   & 3	& -0.33 & 0.11  \\
    1783-33 &   3800  & 0.93 & -0.41   &  1.32    & 75   &  -0.44  & 0.13   & 1	& -0.40 &  ---     \\
\hline
    1978-21 &   3790  & 0.64 & -0.43   &  1.54    &  74  & -0.43   & 0.16   &---&---    &--- \\
    1978-22 &   3700  & 0.55 & -0.37   &  1.50    &  78  & -0.39   & 0.17   & 7	& -0.27 & 0.19 \\
    1978-23 &   3630  & 0.57 & -0.24   &  1.35    &  70  & -0.25   & 0.21   &--- & ---  & ---  \\
    1978-24 &   3750  & 0.62 & -0.30   &  1.40    &  59  & -0.30   & 0.17   & 1	& -0.17 & ---\\
    1978-26 &   3820  & 0.71 & -0.43   &  1.53    &  83  & -0.42   & 0.17   & 1	& -0.28 & ---\\
    1978-28 &   3740  & 0.69 & -0.33   &  1.28    &  85  & -0.33   & 0.18   & 2 & -0.17 & 0.01  \\
    1978-29 &   3750  & 0.71 & -0.44   &  1.58    &  89  & -0.44   & 0.21   & 4	&-0.30  &0.06 \\
    1978-32 &   3700  & 0.73 & -0.40   &  1.39    &  84  & -0.41   & 0.19   & 2	&-0.30  &0.18 \\
    1978-34 &   3900  & 0.83 & -0.32   &  1.49    &  84  & -0.32   & 0.20   &---&  --- & ---\\
    1978-38 &   3840  & 0.81 & -0.43   &  1.59    &  72  & -0.44   & 0.14   & 2	& -0.37 & 0.11\\ 
    1978-42 &   3880  & 0.86 & -0.43   &  1.55    &  92  & -0.43   & 0.18   & 2	& -0.26 & 0.17\\   
\hline
     2173-4 &   3850  & 0.51 & -0.49   &  1.75	  & 97   &  -0.50  & 0.18   & 9	  & -0.49 & 0.11 \\
     2173-5 &   3780  & 0.52 & -0.47   &  1.73	  & 101  &  -0.47  & 0.22   & 3   & -0.37 & 0.13 \\
     2173-6 &   3680  & 0.62 & -0.57   &  1.65	  & 77   &  -0.57  & 0.21   & --- &  ---  & --- \\
     2173-8 &   4060  & 0.83 & -0.57   &  1.72	  & 106  &  -0.58  & 0.14   & 17  & -0.44 & 0.16  \\
    2173-10 &   3880  & 0.91 & -0.42   &  1.65	  & 97   &  -0.41  & 0.18   &  2  & -0.35 & 0.12 \\
\enddata 
\end{deluxetable}

\begin{deluxetable}{ccccccccccccc} 
\footnotesize
\tablecolumns{13} 
\tablewidth{0pc}  
\tablecaption{Chemical abundances, number of measured lines and line-to-line scatter for 
O, Na, Mg and Al.}
\tablehead{ 
\colhead{Star ID}  & \colhead{n}& 
\colhead{[O/Fe]}
 &  \colhead{rms} & 
\colhead{n}  & \colhead{[Na/Fe]} & \colhead{rms}  & \colhead{n} & \colhead{[Mg/Fe]} & \colhead{rms} 
 & \colhead{n} & \colhead{[Al/Fe]} & \colhead{rms}\\ 
 & & \colhead{(dex)}
 &  \colhead{(dex)} &   & \colhead{(dex)} & \colhead{(dex)} & &
  \colhead{(dex)} & \colhead{(dex)}  & & \colhead{(dex)} & \colhead{(dex)} }
\startdata 
 $\log{N}_{\odot}$  &  & 8.79& & & 6.21& & &7.43&  & &6.23 & \\
 \hline
     1651-6 &   2   & -0.12 &  0.05  &   4	 & -0.14 & 0.12   & 3	& +0.16 & 0.13 & 2 & -0.42  & 0.08	  \\
     1651-8 &   2   & -0.07 &  0.08  &   4 	 & -0.41 & 0.07   & 2	& +0.18 & 0.11 & 2 & -0.69  & 0.13	 \\
    1651-10 &   2   & -0.05 &  0.06  &   4	 & -0.16 & 0.05   & 2	& +0.16 & 0.07 & 2 & -0.18  & 0.07	  \\
    1651-12 &   2   & -0.05 &  0.06  &   4  	 & -0.03 & 0.10   & 3	& +0.12 & 0.14 & 2 & -0.28  & 0.06	  \\
    1651-16 &   2   & -0.10 &  0.08  &   4	 & -0.31 & 0.12   & 3	& +0.10 & 0.16 & 2 & -0.59  & 0.02	  \\   
\hline      							   		        		     
    1783-22 &   2   & -0.04 &  0.06  &   4	 & -0.13 & 0.13   & 3	& +0.09 & 0.05 & 2 & -0.45  & 0.12	 \\
    1783-23 &   2   & -0.12 &  0.11  &   4       & -0.12 & 0.09   & 3	& +0.12 & 0.05 & 2 & -0.43  & 0.07	 \\
    1783-29 &   2   &  0.00 &  0.04  &   4       & -0.21 & 0.13   & 4	& +0.09 & 0.09 & 2 & -0.45  & 0.06	      \\
    1783-30 &   2   & -0.12 &  0.07  &   4       & -0.20 & 0.13   & 4	& +0.19 & 0.15 & 2 & -0.36  & 0.08	  \\
    1783-32 &   2   & -0.04 &  0.10  &   4       & +0.03 & 0.08   & 3	& +0.09 & 0.08 & 2 & -0.73  & 0.07	 \\
    1783-33 &   2   & +0.01 &  0.09  &   4       & +0.00 & 0.11   & 4	& +0.17 & 0.14 & 2 & -0.55  & 0.09	 \\    
\hline      
    1978-21 &  2     & -0.03 & 0.06   & 4	  &-0.35 & 0.07   & 4	& +0.17 & 0.16 & 2 & -0.53 & 0.08      \\ 
    1978-22 &  2     & -0.20 & 0.08   & 4	  &-0.05 & 0.13   & 4	& +0.12 & 0.12 & 2 & -0.50 & 0.01	\\
    1978-23 &  2     & -0.15 & 0.07   & 4	  &+0.10 & 0.11   & 4	& +0.21 & 0.09 & 2 & -0.55 & 0.18      \\
    1978-24 &  2     & -0.14 & 0.11   & 4	  &-0.24 & 0.14   & 4	& +0.19 & 0.15 & 2 & -0.41 & 0.11      \\
    1978-26 &  2     & +0.02 & 0.08   & 4	  &-0.25 & 0.11   & 4	& +0.23 & 0.07 & 2 & -0.61 & 0.02      \\   
    1978-28 &  2     & -0.05 & 0.05   & 4	  &-0.06 & 0.15   & 4	& +0.22 & 0.12 & 2 & -0.55 & 0.02	 \\ 
    1978-29 &  2     & -0.06 & 0.08   & 4	  &-0.09 & 0.09   & 4	& +0.22 & 0.10 & 2 & -0.57 &0.05       \\
    1978-32 &  2     & -0.08 & 0.07   & 4	  &-0.19 & 0.09   & 4	& +0.19 & 0.07 & 2 & -0.39 &0.08       \\
    1978-34 &  1     &  0.02 & ---    & 3         &-0.36 & 0.09   & 3	& +0.19 & 0.11 & 2 & -0.56 & 0.05     \\
    1978-38 &  2     & -0.02 & 0.08   & 4	  &-0.24 & 0.13   & 4	& +0.20 & 0.10 & 2 & -0.54 & 0.08	 \\   
    1978-42 &  2     & -0.10 & 0.10   & 4	  &-0.21 & 0.11   & 4	& +0.11 & 0.15 & 2 & -0.55 & 0.02      \\     
\hline    
     2173-4 &  2     & -0.04 & 0.12   & 4	  & +0.27 & 0.11  & 4	&+0.15  & 0.08  & 2 & -0.32   & 0.03	 \\
     2173-5 &  2     & -0.11 & 0.10   & 4         & +0.04 & 0.09  & 4	&+0.07  & 0.13  & 2 & -0.35   & 0.07	 \\
     2173-6 &  2     & -0.07 & 0.07   & 4	  & -0.29 & 0.09  & 3	&+0.09  & 0.05  & 2 & -0.40   & 0.01	  \\
     2173-8 &  2     & -0.08 & 0.07   & 4	  & +0.18 & 0.12  & 4	&+0.15  & 0.08  & 2 & -0.14   & 0.09	   \\
    2173-10 &  2     & -0.05 & 0.08   & 4	  & -0.23 & 0.11  & 4	&+0.04  & 0.16  & 2 & -0.36   & 0.07	   \\
\enddata 		     
\tablecomments{Oxygen abundances are derived 
from spectral synthesis. Sodium abundances are corrected for departures from LTE.}
\end{deluxetable}

\begin{deluxetable}{ccccccccccccc} 
\footnotesize
\tablecolumns{13} 
\tablewidth{0pc}  
\tablecaption{Chemical abundances, number of measured lines and line-to-line scatter for 
Si, Ca, Sc and Ti.}
\tablehead{ 
\colhead{Star ID}  & 
\colhead{n}  & \colhead{[Si/Fe]} & \colhead{rms}  & \colhead{n} & \colhead{[Ca/Fe]} & \colhead{rms} 
 & \colhead{n} & \colhead{[Sc/Fe]II} & \colhead{rms}& \colhead{n}& \colhead{[Ti/Fe]}
 &  \colhead{rms} \\ 
 & & \colhead{(dex)}
 &  \colhead{(dex)} &   & \colhead{(dex)} & \colhead{(dex)} & &
  \colhead{(dex)} & \colhead{(dex)}  & & \colhead{(dex)} & \colhead{(dex)}}
\startdata 
 $\log{N}_{\odot}$  &  & 7.53 & & & 6.27& & &3.13& & &5.00&\\
 \hline
     1651-6  &  4  & -0.12 & 0.10 & 12 &  +0.04  & 0.16  &  5  & -0.06 & 0.12    & 15 & +0.11 & 0.14	   \\
     1651-8  &  3  & +0.00 & 0.04 & 12 &  +0.00  & 0.17  &  5  & -0.11 & 0.10    & 13 & +0.11 & 0.13	  \\
    1651-10  &  3  & -0.01 & 0.10 & 12 & -0.02  & 0.14   &  5  & +0.03 & 0.15    & 11 & +0.09 & 0.13	     \\
    1651-12  &  4  & -0.12 & 0.10 & 11 &  +0.03  & 0.16  &  4  & -0.09 & 0.09    &  9 & +0.04 & 0.06	   \\
    1651-16  &  3  & -0.12 & 0.07 & 11 & -0.07  & 0.13   &  4  & +0.00 & 0.15    & 18 & +0.15 & 0.15	      \\ 
\hline     						   
    1783-22 &  2  & +0.04 & 0.04 & 13 & -0.09  & 0.13    &  5  & -0.06 & 0.16    & 14 & +0.06& 0.16	      \\
    1783-23 &  4  & +0.10 & 0.07 & 13 & -0.13  & 0.12    &  5  & -0.16 & 0.14    & 12 &-0.06 & 0.07	      \\
    1783-29 &  3  & -0.06 & 0.04 & 12 & -0.07  & 0.13    &  5  & -0.04 & 0.16    & 13 & +0.13 & 0.09	     \\
    1783-30 &  3  & -0.04 & 0.06 & 12 & -0.17  & 0.14    &  3  & -0.16 & 0.06    & 13 & +0.03 & 0.15	       \\
    1783-32 &  3  & +0.05 & 0.08 & 13 & -0.17  & 0.09    &  5  & -0.01 & 0.10    & 16 & +0.00 & 0.14	     \\
    1783-33 &  4  & +0.07 & 0.12& 12 & -0.15   & 0.09    &  4  & -0.10 & 0.08    & 12 & +0.01 & 0.14	      \\	   
\hline    						       
    1978-21 &  3  & +0.06 & 0.11 & 13 & -0.14  & 0.13    &  5  & -0.18 & 0.11    & 15 & +0.18 & 0.13	       \\
    1978-22 &  3  & +0.10 & 0.02 & 14 & -0.12  & 0.18    &  4  & -0.15 & 0.06    & 13 & -0.02 & 0.09		\\
    1978-23 &  4  & +0.14 & 0.07 & 14 & -0.10  & 0.15    &  4  & -0.26 & 0.06    & 14 & +0.11  & 0.17		   \\
    1978-24 &  4  & +0.10 & 0.02 & 12 & -0.13  & 0.17    &  4  & -0.06 & 0.13    & 12 & +0.08 & 0.15	     \\
    1978-26 &  4  & +0.05 & 0.12 &  9 & -0.17  & 0.07    &  5  & -0.18 & 0.15    & 15 & +0.16 & 0.15	       \\ 
    1978-28 &  3  & +0.11 & 0.09 & 12 & -0.11  & 0.15    &  5  & -0.14 & 0.15    & 13 & +0.07 & 0.12	    \\
    1978-29 & 4   & +0.14 & 0.08 & 11 & -0.10	&0.17    &  5  & +0.01 & 0.12    & 15 & +0.04 & 0.17	       \\
    1978-32 & 3   & +0.17 & 0.06 & 13 & -0.11  & 0.14    &  3  & -0.23 & 0.08    & 13 & -0.09  & 0.15		\\
    1978-34 & 3   & +0.05 & 0.03 & 12 & -0.08  & 0.17    &  4  & -0.15 & 0.08    & 13 & +0.08 & 0.12	     \\
    1978-38 & 3   & +0.08 & 0.09& 14 & -0.08  & 0.16     &  5  & -0.09 & 0.14    & 16 & +0.00 & 0.19	      \\
    1978-42 & 4   & +0.04 & 0.07 & 8 & -0.18  & 0.08     &  5  & -0.31 & 0.11    &17 & +0.06 & 0.18	     \\   
\hline     
     2173-4 &  4  & +0.07&  0.10 &11  & +0.00 & 0.16     &  4  & -0.11 & 0.10    & 10 & +0.18 & 0.15	\\
     2173-5 &  4  & +0.04 & 0.04 &12  & -0.01 & 0.14     &  5  & -0.19 & 0.13    & 14 & +0.19 & 0.16 \\
     2173-6 &  3  & +0.13 & 0.11 &11  & +0.02 & 0.14     &  4  & -0.14 & 0.09    & 17 & +0.15 & 0.15 \\
     2173-8 &  4  & +0.05 & 0.07 &13  & +0.00 & 0.16	 &  5  & -0.15 & 0.09    & 19 & +0.12 & 0.09 \\
    2173-10 &  4  & +0.07 & 0.10 &12  & -0.09 & 0.16     &  4  & +0.00 & 0.08    & 13 & +0.09 & 0.06 \\
\enddata 
\tablecomments{Scandium abundances include HFS corrections.}
\end{deluxetable}

\begin{deluxetable}{ccccccccccccc}
\footnotesize
\tablecolumns{13} 
\tablewidth{0pc}  
\tablecaption{Chemical abundances, number of measured lines and line-to-line scatter for 
V, Cr, Co and Ni.}
\tablehead{ 
\colhead{Star ID}  & \colhead{n}& \colhead{[V/Fe]}
 &  \colhead{rms} & 
\colhead{n}  & \colhead{[Cr/Fe]} & \colhead{rms}  & \colhead{n} & \colhead{[Co/Fe]} & \colhead{rms} 
   & \colhead{n} & \colhead{[Ni/Fe]} & \colhead{rms} \\ 
 & & \colhead{(dex)}
 &  \colhead{(dex)} &   & \colhead{(dex)} & \colhead{(dex)} & &
  \colhead{(dex)} & \colhead{(dex)}  & & \colhead{(dex)} & \colhead{(dex)}}
\startdata
$\log{N}_{\odot}$  &  & 3.97& & & 5.67& & &4.92&  & &6.28& \\     
\hline
     1651-6   &   8  &+0.09 & 0.08 & 10 & -0.14 & 0.15      & 2  &-0.02 &  0.06    &   23 &+0.00 & 0.16        \\	   
     1651-8   &   8  &+0.08 & 0.13 & 13 &  0.00 & 0.14      & 2  &-0.02 &  0.05    &   24 &-0.04 & 0.13    	 \\
    1651-10   &   6  &+0.05 & 0.06 & 12 & -0.16 & 0.10      & 3  &-0.01 &  0.11    &   23  &+0.07   &0.16  	\\
    1651-12   &   8  &-0.05 & 0.08 & 12 & -0.04 & 0.16      & 3  &-0.06 &  0.02    &   27  &-0.02   & 0.16     \\
    1651-16   &   9  &-0.03 & 0.12 & 15 & -0.08 & 0.15      & 2  &-0.03 &  0.14    &   30  &+0.06   & 0.13     \\  
\hline      
    1783-22   &   7  &+0.14 & 0.11 & 13 & -0.16   & 0.15     &2  &+0.07 &  0.04    &	24  &+0.00   & 0.15       \\
    1783-23   &   7  &-0.13 & 0.15 & 14 & -0.21  & 0.15      &3  &-0.05 &  0.05    &	26  &-0.01   &0.16       \\
    1783-29   &   7  &+0.13 & 0.14 & 13 & -0.05  & 0.16      &3  &+0.05 &  0.09    &	25  &+0.00   &0.12        \\
    1783-30   &   5  &-0.04 & 0.12 & 16 & -0.19  & 0.15      &3  &+0.00 &  0.06    &	25  &-0.05   &0.16       \\
    1783-32   &   7  &+0.05 & 0.15 & 14 & -0.07  & 0.15      &3  &-0.09 &  0.08    &	25 &-0.01   &0.15        \\
    1783-33   &   7  &-0.01 & 0.15 & 14 & -0.05  & 0.16      &3  &+0.05 &  0.10    &	28 &+0.02  & 0.16  	 \\  
\hline       						    	    
    1978-21   &   9  & +0.03 & 0.16 & 15 & -0.12  & 0.16     &3  &-0.12 &  0.15    &	23& +0.02  & 0.14           \\
    1978-22   &   6  &-0.09 & 0.13 & 14 & -0.17  & 0.15      &3  &-0.10 &  0.15    &  26 & +0.08  & 0.15            \\ 
    1978-23   &   5  & +0.16 & 0.07 & 13 & -0.03   & 0.17    &3  & -0.13 &  0.12   &  26 & +0.11  & 0.15              \\ 
    1978-24   &   6  & +0.18 & 0.13 & 14 & -0.13  & 0.16     &2  & +0.09 &  0.05   &  25 & +0.13  & 0.15            \\ 
    1978-26   &   6  & +0.22 & 0.16 & 14 & -0.12  & 0.16     &3  &-0.11 &  0.11    &  25& +0.10  & 0.16             \\
    1978-28   &   5  & +0.15 & 0.08 &  9 & -0.01  & 0.15     &3  & +0.04 &  0.13   &  27 & +0.01  & 0.16            \\    
    1978-29   &   6  & +0.05 & 0.10 & 14 &-0.10  & 0.15      &3  & +0.00 &  0.06   &  23 & +0.04  & 0.10            \\
    1978-32   &   5  & +0.13 & 0.12 & 13 &-0.14   & 0.16     &3  & +0.09 &  0.13   &  24& +0.02  & 0.14              \\
    1978-34   &   6  & +0.00 & 0.17 & 12& -0.20  & 0.15      &2  & +0.00 &  0.05   &  20& +0.11  & 0.16              \\
    1978-38   &   7  &-0.14 & 0.05 & 13 & -0.16   & 0.12     &3  & +0.11 &  0.15   &  31 &-0.05  & 0.15               \\
    1978-42   &   7  &-0.02 & 0.17 & 11 & -0.17  & 0.16      &3  & +0.03 &  0.15   &  31&-0.02  & 0.12	   	    \\
\hline      						    	  
     2173-4  &  7  &  +0.04 & 0.09 & 18 &-0.08 & 0.16        &3  & -0.11&  0.13    &	30  & -0.15  & 0.16    \\   
     2173-5  &  7  &  +0.09 & 0.10 & 12 & +0.03 & 0.12       &3  & +0.03&  0.10    &	27 & -0.05 & 0.1       \\   
     2173-6  &  8  & -0.03 & 0.13 & 11 & -0.05 & 0.17        &3  & +0.01&  0.13    &	16  & -0.07  & 0.13    \\  
     2173-8  & 11  & -0.11 & 0.16 & 17 & -0.11  & 0.14       &3  & -0.10&  0.14    &   35  & -0.04 & 0.13       \\  
    2173-10  &  8  & -0.13 & 0.07 & 12 & -0.06 & 0.14        &4  & -0.20&  0.15    &	32 & -0.05 &  0.15     \\  
\enddata 
\tablecomments{Vanadium and Cobalt abundances include HFS corrections.}
\end{deluxetable}

\begin{deluxetable}{ccccccccccccc} 
\footnotesize
\tablecolumns{13} 
\tablewidth{0pc}  
\tablecaption{Chemical abundances, number of measured lines and line-to-line scatter for 
Y, Zr, Ba and La.}
\tablehead{ 
\colhead{Star ID}  & \colhead{n}& \colhead{[Y/Fe]II}
 &  \colhead{rms} & 
\colhead{n}  & \colhead{[Zr/Fe]} & \colhead{rms}  & \colhead{n} & \colhead{[Ba/Fe]II} & \colhead{rms} 
 & \colhead{n} & \colhead{[La/Fe]II} & \colhead{rms}\\ 
 & & \colhead{(dex)}
 &  \colhead{(dex)} &   & \colhead{(dex)} & \colhead{(dex)} & &
  \colhead{(dex)} & \colhead{(dex)}  & & \colhead{(dex)} & \colhead{(dex)}
 }
\startdata 
 $\log{N}_{\odot}$  &  & 2.22 & & & 2.60& & &2.22& & &1.22&\\
 \hline
     1651-6 &  3  & -0.39 & 0.02  &  3  & -0.53 & 0.08 & 3  & +0.44  & 0.11 & 1 & +0.19 & ---\\
     1651-8 &  2   & -0.27& 0.16   & 3   & -0.41 & 0.13 & 3 & +0.49  & 0.08 & 1 & +0.34 & ---\\
    1651-10 &  3  &-0.51 & 0.11  & 2   & -0.49 & 0.01 & 3   & +0.49  & 0.09 & 1 & +0.26 & ---\\
    1651-12 & 2   & -0.32 & 0.08  & 2   & -0.41 & 0.07& 3   & +0.38  & 0.01 & 1 & +0.12 & ---\\
    1651-16 & 2   & -0.50& 0.12  & 1  & -0.39 & --- & 3     & +0.44  & 0.04 & 1 & +0.11 &--- \\ 
\hline         
    1783-22 & 2   & -0.64& 0.03  & 3   & -0.38 & 0.11 & 3 &  +0.43 & 0.09 & 1 & +0.34& ---   \\
    1783-23 & 2   & -0.40& 0.09  &  3  & -0.57 & 0.07 & 3 & +0.44 & 0.10 & 1 & +0.28& ---   \\
    1783-29 & 2   & -0.48& 0.01  &  3  & -0.57 & 0.07 & 3 & +0.44  & 0.09 & 1 & +0.35 & ---  \\
    1783-30 & 3   & -0.58& 0.14  &  2  & -0.39 & 0.02 & 2 & +0.40 &  0.12& 1 &+0.25  & ---  \\
    1783-32 & 2   & -0.59& 0.02  &  3  & -0.67 & 0.06 & 2 & +0.44  & 0.13 & 1 & +0.37 & ---  \\
    1783-33 & 3   & -0.43 & 0.13  &  2  & -0.64 & 0.04 & 3 &+0.37 & 0.08 & 1 & +0.31 & ---  \\	 
\hline     
    1978-21 &  2  & -0.42& 0.06  &  3  &  -0.29& 0.10 & 3 & +0.48  & 0.11& 1 & +0.31 &--- \\
    1978-22 &  3  & -0.42& 0.08  &  3  & -0.44 & 0.07 & 3 & +0.37  & 0.11& 1& +0.10 & --- \\
    1978-23 &  1  & -0.54& ---  & 2   & -0.26 & 0.07 & 3 & +0.49  & 0.09  & 1 & +0.20 & --- \\
    1978-24 &  2  & -0.41& 0.05  & 2   & -0.52 & 0.06 & 3 & +0.47  & 0.10 &1  & +0.09 & --- \\
    1978-26 &  2  & -0.70& 0.11  & 3   & -0.52 & 0.12 & 2 & +0.37  & 0.04 & 1 &+0.30  & ---  \\ 
    1978-28 &  2  & -0.57& 0.04  & 3   & -0.32 & 0.13 & 3 & +0.52  & 0.10 & 1 & +0.28 &---  \\
    1978-29 &  2  & -0.68 &0.06   & 3   & -0.56 & 0.13 & 3 & +0.34  & 0.13&1  & +0.19 & --- \\
    1978-32 & 2   &-0.50 & 0.02  & 2   & -0.65 & 0.01 & 3 & +0.38  & 0.03 &  1&+0.22  & ---  \\
    1978-34 & 2   & -0.65& 0.11  &  3  & -0.38 & 0.14 & 3 & +0.50  & 0.13 &1&+0.21  & --- \\
    1978-38 & 3   &-0.46 & 0.12  &  3  & -0.48 & 0.05 & 3 & +0.57  & 0.09 & --- & --- & --- \\
    1978-42 & 1   & -0.55& ---  &  3  & -0.55 & 0.08 & 3 & +0.41  & 0.11 & 1&+0.26     &--- \\   
\hline      
     2173-4 & 2   & -0.34& 0.16 &  3  & -0.37 & 0.04 & 3 & +0.44 & 0.08   & 1& +0.17 & ---\\
     2173-5 & 2   & -0.35& 0.05 & 4   & -0.30 & 0.05 & 3 & +0.36 & 0.11   & 1& +0.16 & ---\\
     2173-6 & 2   & -0.35& 0.13 & 3   & -0.35 & 0.07 & 3 & +0.40 & 0.08   & 1& +0.29 & ---\\
     2173-8 & 3   & -0.36& 0.18 & 3   & -0.49 & 0.11 & 3 & +0.47 & 0.09   & 1& +0.19 & ---\\
    2173-10 & 3   & -0.21& 0.14 & 2   & -0.44 & 0.04 & 3 & +0.42 & 0.03   & 1& +0.18 & ---\\
\enddata 
\end{deluxetable}

\begin{deluxetable}{cccccccccc} 
\footnotesize
\tablecolumns{10} 
\tablewidth{0pc}  
\tablecaption{Chemical abundances, number of measured lines and line-to-line scatter for 
Ce, Nd and Eu.}
\tablehead{ 
\colhead{Star ID}  & \colhead{n}& \colhead{[Ce/Fe]II}
 &  \colhead{rms} & 
\colhead{n}  & \colhead{[Nd/Fe]II} & \colhead{rms}  & \colhead{n} & \colhead{[Eu/Fe]II} & \colhead{rms} \\ 
 & & \colhead{(dex)}
 &  \colhead{(dex)} & 
 & \colhead{(dex)} & \colhead{(dex)}  & & \colhead{(dex)} & \colhead{(dex)}
 }
\startdata 
 $\log{N}_{\odot}$  &  & 1.55& & & 1.50& & &0.51& \\
 \hline
     1651-6 &  1   &   -0.03 & ---   &  2    &   +0.41&   0.15 & 1& +0.24 &---  \\  
     1651-8 &  1   &   +0.03 & ---   &  2    &   +0.52&   0.08 & 1& +0.37 &---    \\
    1651-10 &  1   &   +0.16 & ---   &  2    &   +0.30&   0.06 & 1& +0.34 &---    \\
    1651-12 &  1   &   +0.08 & ---   &  3    &   +0.40&   0.03 & 1& +0.44 &---   \\
    1651-16 & ---  &   ---   & ---   &  2    &   +0.22&   0.07 & ---& ---   & --- \\  
\hline        
    1783-22 &  1   &   +0.08 & ---   &  2    &   +0.29&   0.01 & 1&  +0.41& --- \\
    1783-23 &  1   &   +0.00 & ---   &  3    &   +0.22&   0.07 & 1&  +0.37& --- \\
    1783-29 & ---  &   ---   & ---   &  3    &   +0.30&   0.13 & 1&  +0.39& ---  \\
    1783-30 &  1   &   +0.10 & ---   &  2    &   +0.40&   0.12 & 1&  +0.34& --- \\
    1783-32 &  1   &   -0.13 & ---   &  1    &   +0.21&    --- & ---&  ---  & --- \\
    1783-33 &  1   &   -0.08 & ---   &  2    &   +0.49&   0.05 & 1&  +0.57& --- \\  
\hline            			        
    1978-21 &  1   & -0.08   & ---   &  3    &   +0.56&   0.08 & 1&  +0.34& ---      \\
    1978-22 &  1   & +0.05   & ---   &  3    &   +0.44&   0.13 & 1&  +0.51& ---     \\ 
    1978-23 &  1   & -0.02   & ---   &  3    &   +0.39&   0.15 & 1&  +0.34& ---      \\ 
    1978-24 &  --- &  ---    & ---   &  4    &   +0.41&   0.07 & ---&  ---  & ---     \\ 
    1978-26 &  1   & -0.06   & ---   &  3    &   +0.36&   0.07 & 1&  +0.31& ---     \\
    1978-28 &  1   & +0.10   & ---   &  2    &   +0.31&   0.10 & 1&  +0.35& ---     \\    
    1978-29 &  1   & -0.13   & ---   &  3    &   +0.32&   0.14 & 1&  +0.54& ---       \\
    1978-32 &  1   & +0.03   & ---   &  3    &   +0.33&   0.12 & 1&  +0.37& ---      \\
    1978-34 &  --- &  ---    & ---   &  1    &   +0.22&   --- &---& ---   & ---      \\
    1978-38 &  1   & +0.03   & ---   &  2    &   +0.16&   0.02 & 1&  +0.49&---  \\
    1978-42 &  1   & -0.02   & ---   &  2    &   +0.35&   0.08 & ---&  ---  & --- 	 \\
\hline      
     2173-4 &  1   & -0.02   & ---   &  2    &   +0.31 &  0.06 & 1  & +0.37&---  \\   
     2173-5 &  1   & +0.07   & ---   &  3    &   +0.42 &  0.09 & 1  & +0.39&---  \\    
     2173-6 &  1   & +0.00   & ---   &  2    &   +0.21 &  0.07 & 1  & +0.56&--- \\  
     2173-8 &  1   & -0.01   & ---   &  2    &   +0.36 &  0.02 & 1  & +0.49&---  \\  
    2173-10 &  1   & +0.05   & ---   &  2    &   +0.20 &  0.05 & 1  & +0.54&---  \\  
\enddata 
\tablecomments{Europium abundances are derived 
from spectral synthesis. }
\end{deluxetable}

\begin{deluxetable}{ccccccccc} 
\footnotesize
\tablecolumns{9} 
\tablewidth{0pc}  
\tablecaption{Mean abundance ratios for NGC 1651 and NGC 1783.}
\tablehead{ 
& &  \colhead{NGC1651} & & & & \colhead{NGC1783}  & & \\
\colhead{Ratio}  & \colhead{$N_{star}$}& \colhead{Mean}
 &  \colhead{$\sigma_{obs}$} & 
\colhead{$\sigma_{exp}$}  & \colhead{$N_{star}$}& \colhead{Mean}
 &  \colhead{$\sigma_{obs}$} & 
\colhead{$\sigma_{exp}$}  \\
&  & \colhead{(dex)} &  \colhead{(dex)} &  \colhead{(dex)} & 
& \colhead{(dex)} &  \colhead{(dex)} &  \colhead{(dex)}
 }
\startdata 
    $[O/Fe]$	& 5  &  -0.07 & 0.04 & 0.14 & 6  & -0.06   & 0.08 &  0.14   \\
    $[Na/Fe]$	& 5  &  -0.21 & 0.15 & 0.11 & 6  & -0.10   & 0.10 &  0.11\\
    $[Mg/Fe]$	& 5  &  +0.10 & 0.04 & 0.11 & 6	 & +0.12   & 0.04 &  0.10 \\
    $[Al/Fe]$	& 5  &  -0.43 & 0.21 & 0.14 & 6  & -0.49   & 0.13 &  0.14  \\
    $[Si/Fe]$	& 5  &  -0.07 & 0.06 & 0.13 & 6  & +0.03   & 0.06 &  0.12 \\
    $[Ca/Fe]$	& 5  &  +0.00 & 0.04 & 0.13 & 6  & -0.13   & 0.04 &  0.12  \\
    $[Sc/Fe]$II & 5  &  -0.05 & 0.06 & 0.15 & 6  & -0.08   & 0.06 &  0.14 \\
    $[Ti/Fe]$	& 5  &  -0.03 & 0.02 & 0.16 & 6  & +0.03   & 0.06 &  0.14 \\
    $[V/Fe]$	& 5  &  +0.03 & 0.06 & 0.18 & 6  & +0.02   & 0.10 &  0.15 \\
    $[Cr/Fe]$	& 5  &  -0.08 & 0.07 & 0.12 & 6  & -0.12   & 0.07 &  0.12 \\
    $[Fe/H]$	& 5  &  -0.30 & 0.07 & 0.10 & 6  & -0.35   & 0.06 &  0.07  \\
    $[Fe/H]$II  & 5  &  -0.19 & 0.03 & 0.21 & 6  & -0.29   & 0.06 &  0.19 \\
    $[Co/Fe]$	& 5  &  -0.03 & 0.02 & 0.17 & 6  & +0.01   & 0.06 &  0.12 \\
    $[Ni/Fe]$	& 5  &  +0.01 & 0.05 & 0.09 & 6  & -0.01   & 0.02 &  0.07 \\
    $[Y/Fe]$II  & 5  &  -0.40 & 0.11 & 0.16 & 6  & -0.52   & 0.10 &  0.14\\
    $[Zr/Fe]$   & 5  &  -0.45 & 0.06 & 0.22 & 6  & -0.54   & 0.13 &  0.16\\
    $[Ba/Fe]$II & 5  &  +0.45 & 0.04 & 0.13 & 6  & +0.42   & 0.03 &  0.10\\
    $[La/Fe]$II & 5  &  +0.20 & 0.10 & 0.21 & 6  & +0.32   & 0.04 &  0.20\\
    $[Ce/Fe]$II & 4  &  +0.06 &	0.08 & 0.21 & 4  & +0.01   & 0.10 &  0.19\\
    $[Nd/Fe]$II & 5  &  +0.37 & 0.11 & 0.14 & 6  & +0.32   & 0.11 &  0.16\\
    $[Eu/Fe]$II & 4  &  +0.35 & 0.08 & 0.21 & 5  & +0.42   & 0.09 &  0.20\\
\enddata 
\end{deluxetable} 

\clearpage

\begin{deluxetable}{ccccccccc} 
\tablecolumns{9} 
\tablewidth{0pc}  
\tablecaption{Mean abundance ratios for NGC 1978 and NGC 2173.}
\tablehead{ 
&  &\colhead{NGC1978} & & &   & \colhead{NGC2173}  &  \\
\colhead{Ratio}  & \colhead{$N_{star}$}& \colhead{Mean}
 &  \colhead{$\sigma_{obs}$} & 
\colhead{$\sigma_{exp}$}  & \colhead{$N_{star}$}& \colhead{Mean}
 &  \colhead{$\sigma_{obs}$} & 
\colhead{$\sigma_{exp}$} \\ 
&  & \colhead{(dex)} &  \colhead{(dex)} &  \colhead{(dex)} & 
& \colhead{(dex)} &  \colhead{(dex)} &  \colhead{(dex)}}
\startdata 
  $[O/Fe]$    & 11  & -0.11 & 0.08  & 0.14 & 5  & -0.04   & 0.03 & 0.14  \\
  $[Na/Fe]$   & 11  & -0.16 & 0.13  & 0.13 & 5  & +0.01   & 0.25 & 0.11  \\
  $[Mg/Fe]$   & 11  & +0.19 & 0.04  & 0.14 & 5  & +0.10   & 0.05 & 0.10  \\
  $[Al/Fe]$   & 11  & -0.52 & 0.07  & 0.14 & 5  & -0.31   & 0.10 & 0.14  \\
  $[Si/Fe]$   & 11  & +0.09 & 0.04  & 0.12 & 5  & +0.07   & 0.03 & 0.12  \\
  $[Ca/Fe]$   & 11  & -0.11 & 0.05  & 0.15 & 5  & +0.00   & 0.06 & 0.13  \\
  $[Sc/Fe]$II & 11  & -0.17 & 0.09  & 0.15 & 5  & -0.12   & 0.07 & 0.14 \\
  $[Ti/Fe]$   & 11  & +0.08 & 0.07  & 0.16 & 5  & +0.15   & 0.04 & 0.16  \\
  $[V/Fe]$    & 11  & +0.05 & 0.13  & 0.18 & 5  & -0.03   & 0.09 & 0.18 \\
  $[Cr/Fe]$   & 11  & -0.13 & 0.04  & 0.11 & 5  & -0.05   & 0.05 & 0.11  \\
  $[Fe/H]$   & 11  & -0.38 & 0.07  & 0.10 & 5  & -0.51   & 0.07 & 0.09 \\
  $[Fe/H]$II  & 11  & -0.26 & 0.06  & 0.18 & 5  & -0.37   & 0.06 & 0.15 \\
  $[Co/Fe]$   & 11  & -0.01 & 0.09  & 0.14 & 5  & -0.07   & 0.09 & 0.13 \\
  $[Ni/Fe]$   & 11  & +0.05 & 0.06  & 0.09 & 5  & -0.07   & 0.04 & 0.08  \\
  $[Y/Fe]$II  & 11  & -0.54 & 0.11  & 0.14 & 5  & -0.32   & 0.06 &0.15 \\
  $[Zr/Fe]$   & 11  & -0.45 & 0.12  & 0.18 & 5  & -0.39   & 0.07 &0.18 \\
  $[Ba/Fe]$II & 11  & +0.45 & 0.07  & 0.11 & 5  & +0.42   & 0.04 &0.11 \\
  $[La/Fe]$II & 10  & +0.22 & 0.08  & 0.20 & 5  & +0.20   & 0.05 &0.20 \\
  $[Ce/Fe]$II &  9  & -0.01 & 0.07  & 0.20 & 5  & +0.02   & 0.04 &0.21 \\
  $[Nd/Fe]$II & 11  & +0.35 & 0.11  & 0.18 & 5  & +0.30   & 0.09 &0.16 \\
  $[Eu/Fe]$II &  8  & +0.41 & 0.09  & 0.20 & 5  & +0.47   & 0.09 &0.21 \\
\enddata 
\end{deluxetable}

\begin{deluxetable}{ccccccc} 
\tablecolumns{7} 
\tablewidth{0pc}  
\tablecaption{
Comparison between the EWs measured on synthetic 
spectra of Ba lines computed both with and without HFS, for some target 
stars with different atmospheric parameters. HFS components are from 
\citep{pro00}.}
\tablehead{ 
\colhead{Star} & \colhead{$EW_{5853 \mathring{A}}^{NOHFS}$} & \colhead{$EW_{5853 \mathring{A}}^{HFS}$} 
& \colhead{ $EW_{6141 \mathring{A}}^{NOHFS}$} 
& \colhead{$EW_{6141 \mathring{A}}^{HFS}$} & \colhead{$EW_{6496 \mathring{A}}^{NOHFS}$} & \colhead{$EW_{6496 \mathring{A}}^{HFS}$}\\
    &  \colhead{ (m$\mathring{A}$)}  &  \colhead{ (m$\mathring{A}$)}    & \colhead{(m$\mathring{A}$)} &  
    \colhead{ (m$\mathring{A}$)}  & \colhead{(m$\mathring{A}$)} &   \colhead{(m$\mathring{A}$)} }
\startdata 
  1651-6  & 176.3   &  172.6   &   260.5    &  258.3   & 262.0   & 259.5   \\
  1651-12 & 162.9   &  162.2   &   271.4    &  266.9   & 230.9   & 228.5  \\
  1783-22 & 170.7   &  168.8   &   274.0    &  272.2  & 231.0   & 227.8  \\
  1783-29 & 159.5   &  154.0   &   253.9    &  250.4  & 252.1   & 248.5   \\
  1978-22 & 163.4   &  161.3   &   277.4    &  275.3  & 259.5   & 254.7  \\
  1978-42 & 164.1   &  162.7   &   252.9    &  249.2  & 250.4   & 247.8  \\
  2173-6  &  177.6  &  175.3   &   258.4    &  255.0  & 244.3   &  238.9 \\
  2173-8  &  179.1  &  176.4   &   252.7    &  250.5  & 243.0   &  240.3 \\
\enddata 
\label{tab_ba}
\end{deluxetable}

\end{document}